\documentclass[journal]{IEEEtran}

\usepackage{cite}
\usepackage{amsmath,amssymb,amsfonts}
\usepackage{algorithmic}
\usepackage{graphicx}
\usepackage{textcomp}
\usepackage{xcolor}
\usepackage{amsthm}
\usepackage[english]{babel}
\usepackage{multirow}
\usepackage{esint}
\usepackage{algorithm} 
\usepackage{algorithmic} 
\usepackage{caption}
\usepackage{subfigure}
\usepackage{lipsum}
\usepackage{multicol}
\usepackage{comment}
\usepackage{breqn}
\usepackage{float}

\begin{document}

\title{A Joint Intensity-Neuromorphic Event Imaging System for Resource Constrained Devices}

\author{Srutarshi~Banerjee,
        Henry~H.~Chopp,
        Jianping~Zhang,~Zihao~W.~Wang,~Oliver~Cossairt, and
        Aggelos~Katsaggelos
        % <-this % stops a space
\thanks{Manuscript received on May 24, 2021. This work was supported by a DARPA Contract No. HR0011-17-2-0044.}
\thanks{Srutarshi Banerjee, Henry H. Chopp and A. K. Katsaggelos are with the Department of Electrical and Computer Engineering, Northwestern University, Evanston, IL 60208, USA. (email: srutarshibanerjee2022@u.northwestern.edu, henrychopp2017@u.northwestern.edu, a-katsaggelos@northwestern.edu).}% <-this % stops a space
\thanks{Jianping Zhang and Zihao W. Wang recently graduated from Department of Computer Science, Northwestern University, Evanston, IL 60208, USA. (email: jianpingzhang2018@u.northwestern.edu, zwinswang@gmail.com).}% <-this % stops a space
\thanks{Oliver Cossairt is with Department of Computer Science, Department of Electrical and Computer Engineering, Northwestern University, Evanston, IL 60208, USA. (email: oliver.cossairt@northwestern.edu).}
\thanks{This work has been submitted to the IEEE for possible publication. Copyright may be transferred without notice, after which this version may no longer be accessible.}}

\markboth{Submitted to IEEE Transactions on Neural Networks and Learning Systems, 25 May 2021}%
{Shell \MakeLowercase{\textit{et al.}}: Bare Demo of IEEEtran.cls for IEEE Journals}

\maketitle

\begin{abstract}
We present a novel adaptive multi-modal intensity-event algorithm to optimize an overall objective of object tracking under bit rate constraints for a host-chip architecture. The chip is a computationally resource constrained device acquiring high resolution intensity frames and events, while the host is capable of performing computationally expensive tasks. We develop a joint intensity-neuromorphic event rate-distortion compression framework with a quadtree (QT) based compression of intensity and events scheme. The data acquisition on the chip is driven by the presence of objects of interest in the scene as detected by an object detector. The most informative intensity and event data are communicated to the host under rate constraints, so that the best possible tracking performance is obtained. The detection and tracking of objects in the scene are done on the distorted data at the host. Intensity and events are jointly used in a fusion framework to enhance the quality of the distorted images, so as to improve the object detection and tracking performance. The performance assessment of the overall system is done in terms of the multiple object tracking accuracy (MOTA) score. Compared to using intensity modality only, there is an improvement in MOTA using both these modalities in different scenarios.   
\end{abstract}


% \fi
\IEEEpeerreviewmaketitle

\section{Introduction}
\IEEEPARstart{T}{his} work focuses on the problem of optimal information extraction for a particular task from multiple modalities using high resolution imaging sensors, specifically RGB and event sensors. For high resolution sensors, the data generated result in high bit rates (often $> 1$ Gbits/s). Primary challenges for such systems are the storage and transmission of the data over communication networks. In practice, the available data bandwidth is often limited and time varying, due to various factors, such as, lossy transmission media, and network traffic. This is further complicated due to transmission of data in multiple modalities, such as RGB-infrared, RGB-events.

Event cameras are novel sensors which capture visual information in a drastically different form. Instead of outputting the intensity signals as in traditional cameras, they compare the difference between the current log intensity state and the previous state and fire an event when it exceeds the firing positive or negative threshold. Compared to traditional cameras, event sensing provides several benefits such as low latency operation of individual pixels, high dynamic range, reduce redundant capturing of static scenes, low power consumption.

The issue of selecting task specific appropriate data from either modality is critical. This problem can be framed as a rate-distortion optimization one in multiple modalities prior to transmitting data. While there can be enough computational power on the chip (remote device on the field) for processing data, followed by transmitting only vital information, we develop our framework based on a host-chip architecture, where the chip is resource constrained to perform limited computations. It may be argued that multiple modalities represent redundant information in the scene. However, this is not the case as different imaging modalities capture data in different methods and forms and each modality provides complimentary information to the other modalities. Additionally, processing algorithms rely on joint modalities to improve upon their performance. For instance several such works have been done in the RGB-IR domain: \cite{takumi2017multispectral}, \cite{saha2019multispectral}, and in the intensity-event domain: \cite{zihao2018_EVFlowNet}, \cite{wang2019event}, just to refer a few. In this work we focus on the intensity and neuromorphic event modalities which is gaining popularity in recent years. Several neuromorphic event cameras such as DAVIS346 \cite{posch2010qvga}, CeleX-V \cite{Celex-V} provide intensity frames and events. 

Rate-distortion or resource allocation optimization has been a fundamental approach, addressed mostly for video/image compression and transmission over the last few decades. However, in recent years, with the advancement of other imaging modalities, the fundamental rate-distortion formulation needs to be effectively addressed especially for the multi-modality imaging framework, with intensity - event modality being one such multi-modal system. To the best of authors' knowledge, there has been no work addressing the joint rate-distortion optimization for intensity-event modalities. Moreoever, in this work, we develop the host-chip framework to optimize the rate-distortion equation together with the reconstruction algorithms for object detection and tracking based on both these modalities. While compression of these data can be done in several ways, in this work we use a QT decomposition of the intensity-event volume. The QT structure has been used as QT based block compression schemes fit into the most popular encoding schemes, such as, VVC \cite{VVC_doc}, HEVC \cite{hevc}. The architecture is based on a host-chip model, with the chip being the computationally constrained imager on the field, while the host being a server with high computational power. While this system has the flexibility of optimizing any end to end task, in this work, object tracking is the goal. In our previous work, \cite{eusipco} and \cite{ReImagine_Sys_1}, we solved the rate-distortion problem for only the intensity modality in discrete spatio-temporal volume using QT compression optimized (on the chip), together with a reconstruction algorithm for object detection and tracking. In this work, we jointly solve the rate-distortion problem for the intensity and asynchronous event modalities in a continuous spatio-temporal volume incorporating QT compression and lossy event compression strategy (on the chip), together with a reconstruction algorithm for object detection and tracking. While in our previous work, the host performs object detection and tracking in reconstructed intensity frames, in this work, the host reconstructs intensity frames with the aid of events, followed by object detection and tracking separately in intensity and event modalities, before finally fusing them. To the best of authors' knowledge, the joint rate-distortion optimization of intensity-event modalities for object detection and tracking has not been studied in the literature. The main contributions of the paper are as follows:
\begin{itemize}
    \item Development of a host-chip architecture for optimal information extraction using intensity and event modalities in a computationally resource constrained chip and bandwidth constrained communication channel. 
    \item Development of resource allocation and a joint intensity and event rate-distortion framework on chip.
    \item Development of detection and tracking in intensity and event modalities separately with a late fusion model.
    \item Task-specific processing of distorted intensity and events for object tracking.
\end{itemize}

\section {Related Work}
 Conventional image and video coding techniques have successfully evolved over the past several decades. This has resulted in a number of video standards, some of the latest ones being VVC, HEVC \cite{VVC_doc}, \cite{hevc}. These sophisticated standards take into consideration rate constraints, quality constraints among other factors. Compression schemes specifically for events have evolved in recent years \cite{bi2018spike}, \cite{khan2020time} exploiting spatial and temporal correlation among events. Our work on event compression \cite{banerjee2020lossy} exploited spatio-temporal redundancy along with Poisson Disk Sampling \cite{bridson2007fast} in order to compress events in a lossy manner using a QT optimized scheme and intensity modality, under rate constraints.

%\emph{Region based Rate Control Optimization}
Most of the work in the literature on rate-distortion optimization for video / images do not take into account regions of interest (RoI) while performing rate limited compression. One of the early works on region-based rate control for H.264 is described in \cite{Rate_Region_Control}. The authors used a region-based rate control scheme for macroblocks and grouped regions of similar characteristics as same region for treating them as a basic unit for rate control. Similar works on region-classification-based rate control for Coding Tree Units (CTUs) in I-frames to improve its reconstruction quality for suppressing flicker artifacts has been done in \cite{Region_class_I_frames}. Region-based intra-frame rate-control scheme to improve the objective quality and reduce PSNR fluctuations among CTUs \cite{Region_based_CTU} has been done, while moving regions has been used as the RoI for identifying the depth level of CTU \cite{MR_RoI_CTU}, and other works \cite{hu2017region}, \cite{wang2018region}.

Further progress has been made in the RoI aware rate-control, where a higher bit rate is allocated to regions with human faces \cite{li2016region} and combination of human faces with CTU level \cite{meddeb2014region} and using human faces and tile-based rate control \cite{meddeb2015icip}. Work has been done in attention region based rate control for 3DVC depth map coding based on regions classified as foreground, edges of objects and dynamic regions \cite{lee2016attention}. A content-aware rate control scheme for HEVC has been done on static and dynamic saliency detection using deep convolutional network for extracting static saliency map \cite{sun2020content}. Work has been done on preserving scale-invariant features such as SIFT/SURF \cite{SURF_preserve}. The rate control algorithm of HEVC is based on RoI using an improved Itti algorithm \cite{song2020optimized}. Rate control has also been done via adjustment of $\lambda$ \cite{rate_control_lambda} and in other works. Work on the joint optimization of multiple modalities is scarce in literature. For point clouds, joint bit allocation problem for geometry and color in the point clouds has been solved using a model based approach using interior point method \cite{liu2020model}. In \cite{2020_icmew}, a 2 step approach was used for encoding geometry and color video sequences in a video based point cloud compression (V-PCC) standard.

Object detection and tracking has been one of topics studied widely which has been accelerated with the advances in Deep Learning. Such intensity based object detector are presented in \cite{ren2015faster}, \cite{lin2017feature}, \cite{bochkovskiy2020yolov4}. Several works on intensity based object trackers using both deep learning and non deep learning based approaches such as \cite{yang2016temporal}, \cite{bewley2016simple}, \cite{voigtlaender2019mots}. Event based processing has also been done in recent years to solve classical as well as novel problems in computer vision and robotics. For instance event only based object detection has been done in \cite{cannici_YOLE}, \cite{mitrokhin2018event}, \cite{gehrig2019end}. Object tracking based on events have been done, for example, in \cite{zhu2017event}, \cite{mitrokhin2018event}, \cite{chen2019asynchronous}. Deblurring of objects using events has been done in \cite{lin2020learning}, \cite{zhang2020hybrid} and other algorithms based on events have been developed.

Clearly, there has not been any work in the literature (to the best of authors' knowledge) on optimal information extraction based on joint intensity - event modalities constrained on computationally constrained chip devices at low bit rates in order to optimize end-to-end task-specific objectives, such as object tracking. We address this vital issue in this paper.

\section{Multi modal Host-Chip Architecture}
\begin{figure}[htbp]
\centerline{\includegraphics [width=0.85\linewidth]{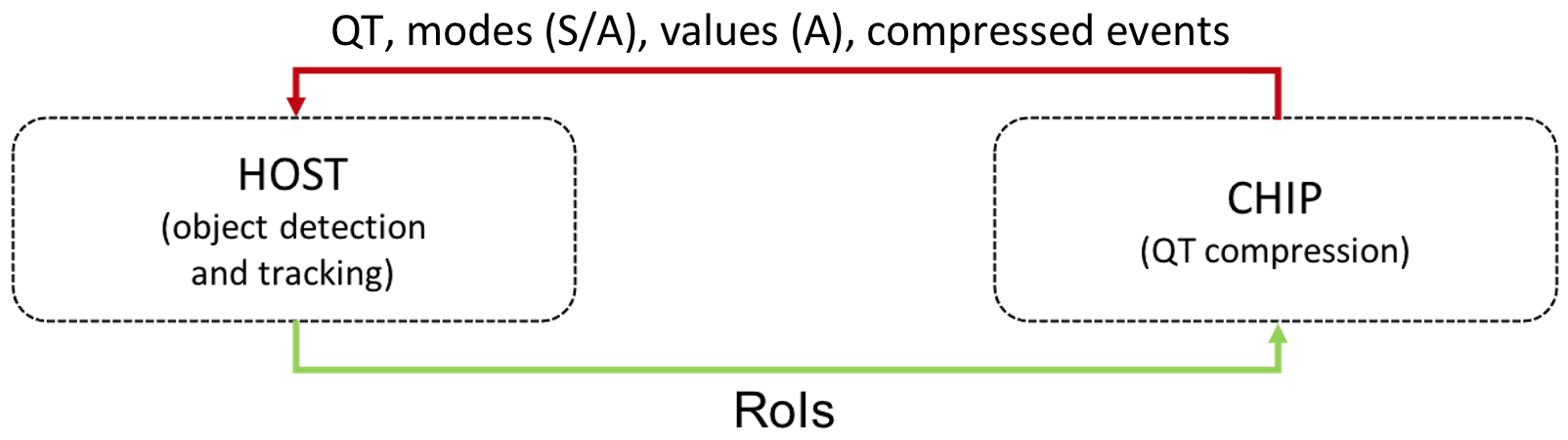}}
\caption{Host-Chip Architecture (S: Skip, A: Acquire)}
\label{fig_overview_pic}
\end{figure}

The system architecture consists of a host and a chip for object detection and tracking in a multi-modal framework. The multiple modalities used in this work are the grayscale intensity and neuromorphic events. While grayscale intensity frames are generated at a fixed frame rate, the events are inherently asynchronous in nature, with each event represented as $(x,y,t,p)$, where $(x,y)$ is the position, t the timestamp and p the polarity. We develop a framework as a prediction-correction feedback system (shown in Fig. \ref{fig_overview_pic}) which is able to work with synchronous (frame-based) and asynchronous data under the constraints of a limited bandwidth channel capacity $B$ between the host and the chip, remotely deployed in the field. The host predicts the Regions of Interest (RoIs) in a frame and updates its prediction based on the data received from the chip. The chip acquires high resolution intensity frames at every time $t$ and corresponding events in a scene asynchronously. The communication bandwidth between the chip and host is limited. In such a scenario, it is not possible for the chip to transmit all the captured intensity and event data to the host. The intensity and event data are compressed using a QT structure in an optimal rate-distortion sense. For the intensity frames, the QT, modes of leaves (skip or acquire) and pixel values corresponding to the acquire mode are sent to the host. On the other hand, the asynchronous events are first quantized in time and then sampled as per the Poisson Disk Sampling method \cite{bridson2007fast}. The sampled events are then compressed as per the QT blocks before being transmitted to the host. The host reconstructs the distorted intensity frames and events based on the data sent by the chip and the previous reconstructed intensity frame. The reconstructed intensity frames and events are used for enhancing the reconstructed frame, object detection and tracking in intensity and event modalities in order to extract the RoIs in each modality. The RoIs from each modality are fused to predict the overall RoIs at the next time instance $t+1$, which are then sent to the chip. Figure \ref{fig_overview_pic} shows the predictive-corrective feedback loop between the chip and the host.

\begin{figure}[htbp]
\centerline{\includegraphics [width=\linewidth]{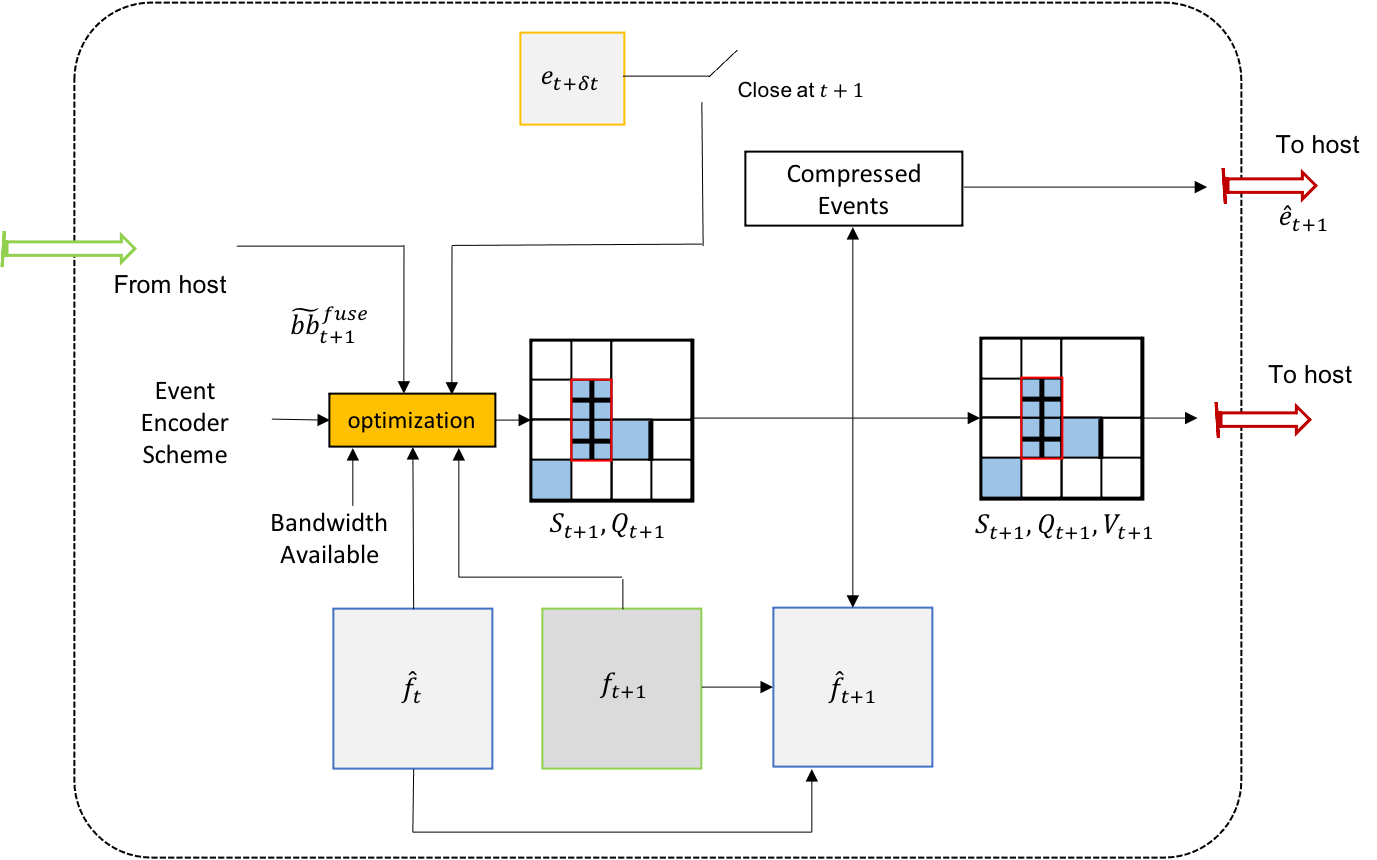}}
\caption{Chip Architecture}
\label{fig_1}
\end{figure}
\subsection{Chip Computation}
Figure \ref{fig_1} shows the chip architecture. The undistorted grayscale frame at $t+1$, $f_{t+1}$ and events between time $t$ and $t+1$, $e_{t+1}$ captured by the chip occupy high communication bandwidth. Hence, the data need to be compressed before transmitting them from the chip to the host. A QT based compression of the intensity frame and events is used in this work. The optimal QT structure, $S_{t+1}$, is obtained by jointly solving the rate-distortion equation for both intensity and events by taking into account the priority regions $\widetilde{bb}^{fuse}_{t+1}$ as well. The priority regions $\widetilde{bb}^{fuse}_{t+1}$ is computed on the host based on the intensity and event data sent from the chip. The host computed $\widetilde{bb}^{fuse}_{t+1}$ is sent to the chip before every time instance $t+1$. The rate-distortion optimization is solved using dynamic programming based Viterbi optimization. For the intensity modality, the Viterbi optimization also generates skip-acquire modes $Q_{t+1}$, corresponding to the QT leaves. For the acquire mode in the QT, the superpixel values $V_{t+1}$ for the leaves of QT are obtained from $f_{t+1}$. The events are sampled and compressed based on the QT using Huffman and run length encoding, generating compressed coded events $\hat{e}_{t+1}$. $S_{t+1}$, $Q_{t+1}$, $V_{t+1}$ along with $\hat{e}_{t+1}$ are sent from the chip to the host. The distorted frame $\hat{f}_t$ at time $t$ stored in the chip, is used for computation at time $t+1$.

%\subsubsection{Joint Optimization of multiple modalities}
\subsubsection{Operational Rate Distortion Optimization based on multiple modalities}
\label{math_joint_model}
The communication bandwidth between chip and host limits the amount of data to be transferred between them. Moreover, due to the presence of image and events modalities, the optimal allocation of bits among these two modalities is very critical. In our previous work \cite{eusipco} and \cite{ReImagine_Sys_1}, we used dynamic programming based Viterbi optimization for controlling the trade-off between the frame distortion and bit rate for the intensity modality only. In this work, we extend the Viterbi optimization framework for performing rate-distortion optimization on both intensity and event modalities. Additionally, the algorithm determines the optimal bit rate allocation between intensity and events in order to minimize the total distortion. The number of bits allocated to intensity and events depends on the compression strategies applied to them. For a pixel intensity value corresponding to the acquire mode, $8$ bits are used to represent it. On the other hand, the events are first sampled using the Poisson disk sampling method in the QT along with Huffman and run length encoding, similar to our work in \cite{banerjee2020lossy}. In it, the QT is result from the rate-distortion optimization of intensity frames and then event compression is performed based on different QT block sizes. In this work, the QT has been obtained by joint rate-distortion optimization on both intensity and event modalities considering the event compression strategy in each QT block as in \cite{banerjee2020lossy}.

The event timestamps are truncated to place the events spatially into $N$ temporal bins, ($N$ chosen to be equal to $4$ in the experiments). One event volume with $N$ temporal bins is considered in the QT blocks with the Poisson disk sampling technique having Poisson Disk Radius (PDR), $r$, to sample events in the QT blocks. In this compression scheme, smaller $r$ is chosen for smaller blocks and larger $r$ for larger blocks, as we prioritize smaller blocks over larger blocks, more events are sampled relatively from smaller blocks. We apply PDR for QT blocks of size $\geq 4$. For QT blocks of size smaller than $4$, we consider all events as important and sample them all. In general it is possible to choose different $r$ candidate values for each QT block and then optimize over the entire QT and $r$. However, the larger the number of candidate $r$ values, the larger the required number of computations. In general, we can have $M_{\tau}$ values for the PDR, $r_{0}, r_{1}, ..., r_{M_{\tau}-1}$ to be optimized in addition to $N_\tau$ leaves of the QT, where $M_{\tau} \leq N_{\tau}$. The total distortion ($D$) is the sum of the intensity frame distortion ($D_i$) and event distortion ($D_e$), over each leaf $x$ of the QT and PDR value $r$, 
\begin{align}
\label{eqn:eqlabel1}
    D(x, r) = D_i(x) + D_e(x, r),
\end{align}
where, $x \in \{x{_0}, x{_1}, ..., x_{N_{\tau}-1}\}$ and $r \in \{r{_0}, r{_1}, ..., r_{M_{\tau}-1}\}$.

Similarly, the total rate ($Ra$) is the sum of the intensity frame rate ($R_i$) and event rate ($R_e$), over each leaf $x$ of the QT and PDR value $r$, that is,
\begin{align}
\label{eqn:eqlabel2}
    Ra(x, r) = R_i(x) + R_e(x, r),
\end{align}
Thus given a maximum bit rate $R_{max}$, we formulate the following rate-distortion optimization problem
\begin{align}
\label{eqn:eqlabel3}
   \arg\min_{\textbf{x, r}} & {~D_{i}(\textbf{x}) + D_{e}(\textbf{x, r})}, \\ 
   %\argmin_{\bx} & {~D(\bx)}, \\
   \text{s. t. } & {R_i(\textbf{x}) + R_e(\textbf{x, r}) \leq R_{max}}. \nonumber
\end{align}
The constrained discrete optimization problem of Eqn. \ref{eqn:eqlabel3} is solved using Lagrangian relaxation, leading to solutions in the convex hull of the rate-distortion curve for single modality \cite{schuster1997video}, \cite{schuster1998optimal}. For dual modality, the Lagrangian cost function is, 
\begin{align}
\label{eqn:eqlabel4}
    J_{\lambda}(\textbf{x, r}) = D_i(\textbf{x}) + D_e(\textbf{x, r}) + \lambda \{R_i(\textbf{x}) + R_e(\textbf{x, r})\},
\end{align}
where $\lambda \geq 0$ is a Lagrangian multiplier.
Equation \ref{eqn:eqlabel4} can be rewritten as, 
\begin{align}
\begin{split}
\label{eqn:eqlabel5}
J_{\lambda}(x_{0}, ..., x_{N_{\tau}-1}, r_{0}, ..., r_{M_{\tau}-1}) = D_i(x_{0}, x_{1}, ..., x_{N_{\tau}-1})\\
+ D_e(x_{0}, ..., x_{N_{\tau}-1}, r_{0}, ..., r_{M_{\tau}-1}) 
+ \lambda \{R_i(x_{0}, ..., x_{N_{\tau}-1})\\
+ R_e(x_{0}, ..., x_{N_{\tau}-1}, r_{0}, ..., r_{M_{\tau}-1})\},
\end{split}
\end{align}
which can be written as the following minimization problem, 
\begin{align}
\label{eqn:eqlabel6}
\arg\min_{(x_{0}, ..., x_{N_{\tau}-1}, r_{0}, ..., r_{M_{\tau}-1})} & 
G(x_{0}, ..., x_{N_{\tau}-1}, r_{0}, ..., r_{M_{\tau}-1}).
\end{align}
The goal is to solve Eqn. \ref{eqn:eqlabel6} using Dynamic Programming (DP) to find the optimal state sequence $x^*_0, x^*_1, ..., x^*_{N_{\tau}-1}$, over leaves of QT and $r^*_0, r^*_1, ..., r^*_{M_{\tau}-1}$, over PDR at each leaf. The PDR is optimized over $M_{\tau}$ leaves out of $N_{\tau}$ in the QT, where $M_{\tau} \leq N_{\tau}$. Considering $g^*_k(x_k, r_k)$ as the minimum cost up to epoch $k$, with $n_0$ is the root of the QT, and, $4^{N-{n_0}}-1 \geq k \geq 0$, we can write,
\begin{align}
\label{eqn:eqlabel7}
g^*_k({x_k, r_k}) = \min_{(x_{0}, x_{1}, ..., x_{k}, r_{0}, r_{1}, ..., r_{k})} & \sum_{j=0}^{k} g(x_j, r_j),
\end{align}
where 
\begin{align}
\begin{split}
\label{eqn:eqlabel8}
g(x_j, r_j) = j_{\lambda}(x_j, r_j) = 
d(x_{j}, r_{j}) + (\lambda \times r_{a}(x_j, r_j))
\end{split}
\end{align}
is the Lagrangian cost function for the $j$ th block with $d(x_{j}, r_{j})$ and $r_{a}(x_j, r_j)$ being the distortion and rate respectively.
Now, 
\begin{dmath}
\label{eqn:eqlabel9}
g^*_{k+1}({x_{k+1}, r_{k+1}}) = \min_{(x_{0}, ..., x_{k+1}, r_{0}, ..., r_{k+1})}  \sum_{j=0}^{k+1} g(x_j, r_j) \\
=\min_{x_{k+1}, r_{k+1}}\Big\{{\min_{x_{0}, ..., x_{k}, r_{0}, ..., r_{k}}}  \left[\sum_{j=0}^{k} g(x_j, r_j)\right]
+ g(x_{k+1}, r_{k+1})\Big\}
\end{dmath}
which results in the DP recursion formula shown in Eqn. \ref{eqn:eqlabel10} with $g^*_{k+1}(x_{k+1}, r_{k+1})$, the minimum cost up to epoch $k+1$, 
\begin{dmath}
\label{eqn:eqlabel10}
g^*_{k+1}(x_{k+1}, r_{k+1}) = \min_{x_{k+1}, r_{k+1}}
\{g^*_{k}(x_k, r_k) + g(x_{k+1}, r_{k+1}) \}
= g^*_{k}(x_k, r_k) + \min_{x_{k+1}, r_{k+1}} g(x_{k+1}, r_{k+1})
\end{dmath}

The DP forward recursion algorithm can be used to find the optimal state sequence. In every epoch (out of $4^{(N-n_{0})}$) in the Viterbi algorithm, the shortest path is found over the set of all admissible nodes of the previous epoch $k-1$ to every node in the set of admissible nodes in the current epoch $k$ which are referred as "from" set $F_{l,i}$ "to" set $T_{l,i}$, where $l$ and $i$ are level and block of the QT. In each block there are $p$ values of PDRs to chose from in order to optimize over $r$. The sequence of initialization, recursion, termination and backtracking for the forward DP algorithm \cite{schuster1997video}, \cite{schuster1998optimal} has been followed in order to obtain the optimal state sequence $x^*_{0}, x^*_{1}, ..., x^*_{N_{\tau}-1}$, $r^*_{0}, r^*_{1}, ..., r^*_{M_{\tau}-1}$.

For a given $\lambda$, $g(x_{k+1}, r_{k+1})$ in Eqn. \ref{eqn:eqlabel10} can be defined as:
\begin{dmath}
\label{eqn:eqlabel11}
g(x_{k+1}, r_{k+1}) = d_i(x_{k+1}) + d_e(x_{k+1}, r_{k+1}) + \lambda \{r_i(x_{k+1}) + r_e(x_{k+1}, r_{k+1})\}
\end{dmath}
\begin{figure*}[h!]
\begin{center}
\includegraphics[height=170 pt]{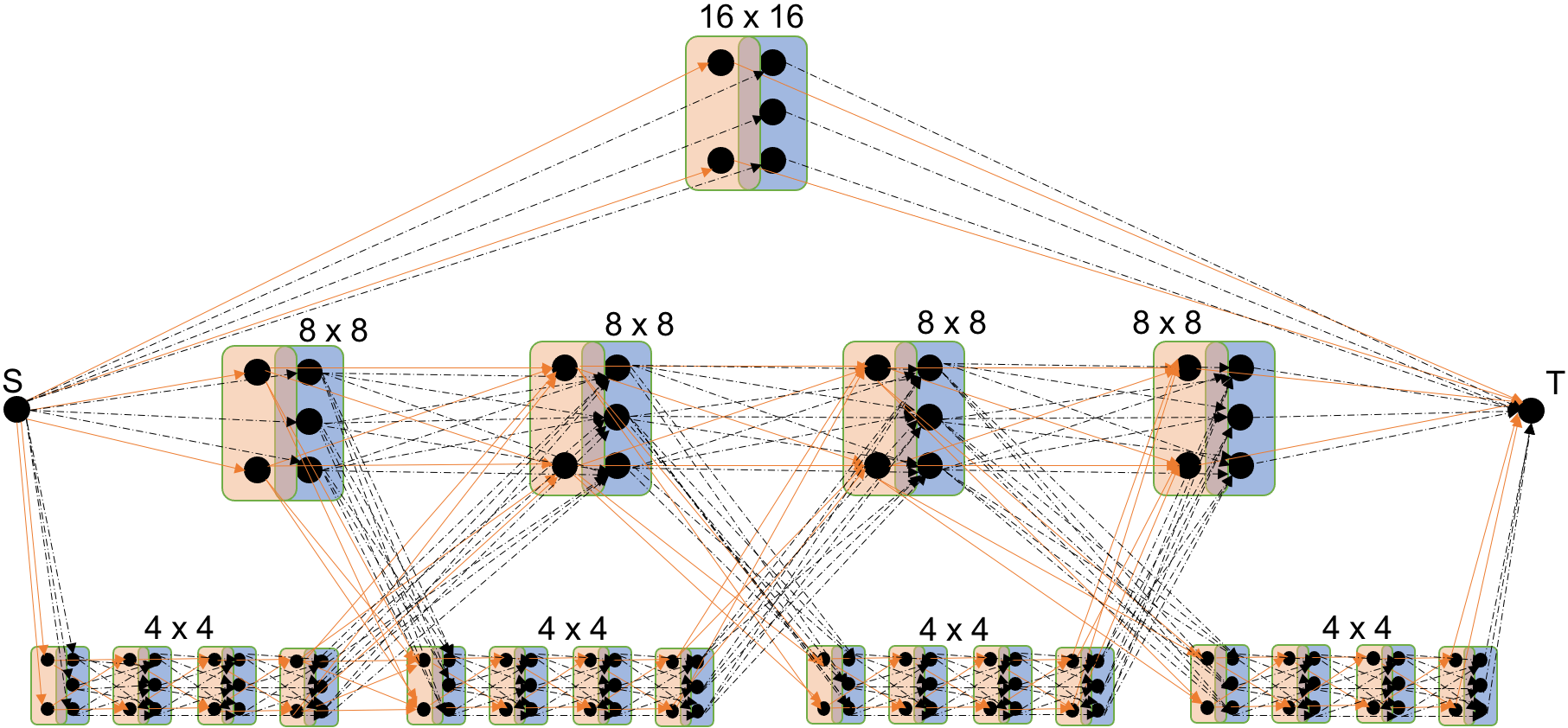}
\end{center}
   \caption{Multilevel trellis with possible transitions for $N = 4$  and $n_0 = 2$.}
\label{fig: Trellis_transitions}
\end{figure*}
Figure \ref{fig: Trellis_transitions} represents the multi-level trellis for a $16 \times 16$ block of image (event) frame $(N=4)$, with a QT segmentation developed down to level 2 $(n_0 = 4 \times 4)$. In actual implementation, for a $512 \times 512$ frame, $N = 9$ and $n_0 = 0$. Each QT node has intensity QT nodes and event QT nodes, denoted by pink and blue color in Fig. \ref{fig: Trellis_transitions}, respectively. The transitions between the intensity QT nodes are shown in pink arrows while transition between the event QT nodes are shown in black dashed arrows. The number of nodes in the QT denotes the number of admissible states in the QT. Although there can be several nodes in the intensity QT and event QT nodes, for simplicity only $2$ nodes are shown for the intensity QT (denoting skip/acquire modes) and $3$ nodes in the event QT (denoting $3$ candidate PDR values) at each node. The intensity rate $r_i(x_{k+1})$ for node $x_{k+1}$ is further sub-divided as the sum of $r_{seg}(x_{k+1})$, $r_{mode}(x_{k+1})$ and $r_v(x_{k+1})$ which are the bit rate allocated for the segmentation, skip/acquire modes and values for the intensity pixels in the acquire mode, respectively.  

The distortion as described in Eqn. \ref{eqn:eqlabel1} can be described as the weighted average of the distortions at each leaf due to the intensity and event. In our previous work, \cite{ReImagine_Sys_1}, the distortion was used as the weighted distortion normalized over the area $A_i$ for each RoI. In this work, we add the weighted distortion on the events (with the weight parameter $w_e$) to the intensity distortion, that is,
\begin{dmath}
\label{eqn:eqlabel12}
D_{Tot} = \sum_{i \in \Omega} \frac {w_{i} D_{i}(x_i)}{A_i} + w_{e}D_{e}(x_i, r_i),
\end{dmath}
where, $\Omega$ is the set of differently weighted regions.
The distortion for the intensity differs for the skip / acquire modalities in each leaf node as described in \cite{ReImagine_Sys_1}. The distortion for the events in each leaf node is described by Eqn. \ref{eqn:eqlabel13}. $E_{org}(i,j)$ and $E_{dist}(i,j)$ are the aggregated event counts in pixel $(i,j)$ along the temporal axis for the original (undistorted) and distorted events, respectively. $N_{bl,events}$ are the QT blocks containing events. The aggregation step is done without accounting for the polarity of the events. Thus, event count takes into account both positive and negative events.
\begin{dmath}
\label{eqn:eqlabel13}
D_{e}(x_i, r_i) = \sum_{(i,j) \in N_{bl,events}} (E_{org}(i,j) - E_{dist}(i,j))
\end{dmath}
In order to operate the system at fixed bit rate (within certain tolerance), the $\lambda$ value in Eqn. \ref{eqn:eqlabel4} is adjusted in each frame. The optimal $\lambda^*$ is computed by a convex search in the Bezier curve \cite{schuster1998optimal}, which accelerates convergence.

\subsection{Host Computation}
\begin{figure}[htbp]
\centerline{\includegraphics [width=\linewidth]{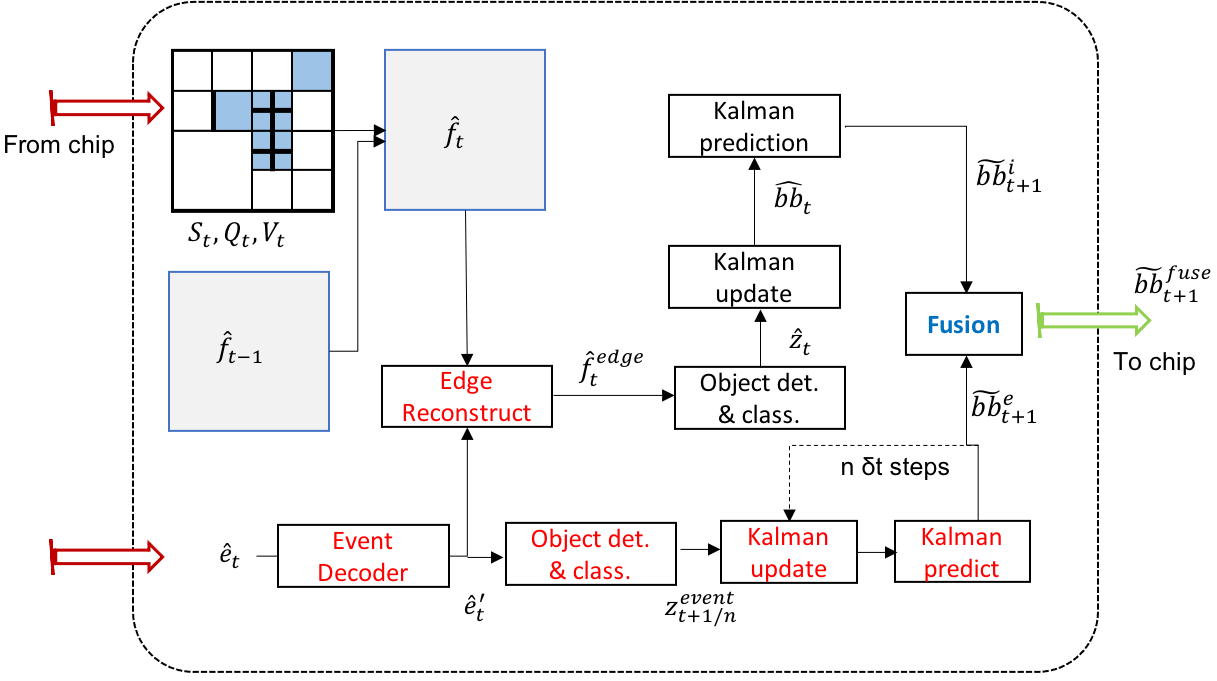}}
\caption{Host Architecture}
\label{fig_2}
\end{figure}

Figure \ref{fig_2} shows the computation on the host. The host receives from the chip QT, skip/acquire modes and values corresponding to the acquire modes at time $t$, denoted as $S_t$, $Q_t$ and $V_t$, respectively, along with compressed events $\hat{e}_{t}$. The reconstructed frame $\hat{f}_{t}$ is generated from $S_t$, $Q_t$, $V_t$ and the reconstructed frame $\hat{f}_{t-1}$ at $t-1$. The coded events $\hat{e}_{t}$ are decoded as ${\hat{e}}'_t$ which is an event-frame representation, before further processing. ${\hat{e}}'_t$ and $\hat{f}_t$ are used for refining the edges in the reconstructed frame $\hat{f}_t$ to create the enhanced frame $\hat{f}^{edge}_t$, which is then used for object detection and classification generating bounding boxes which are then fed to a Kalman filter based tracker. The events ${\hat{e}}'_t$ are additionally used to perform object detection and classification $n$ times between times $t$ and $t+1$ before feeding them as observations to the Kalman filter-based tracker for generating bounding boxes every $n$ timesteps between $t$ and $t+1$. At time $t+1$, the predicted bounding boxes generated from the event based detections are fused with those from the intensity based detections to generate fused bounding boxes $\widetilde{bb}^{fuse}_{t+1}$ which are then sent to the chip as priority regions for the Viterbi optimization. These steps are shown in Fig. \ref{fig_2}.

\subsubsection{Intensity based Processing}
\label{Intensity based Processing}
\begin{figure}[htbp]
\centerline{\includegraphics[width=0.9\linewidth, height=5cm]{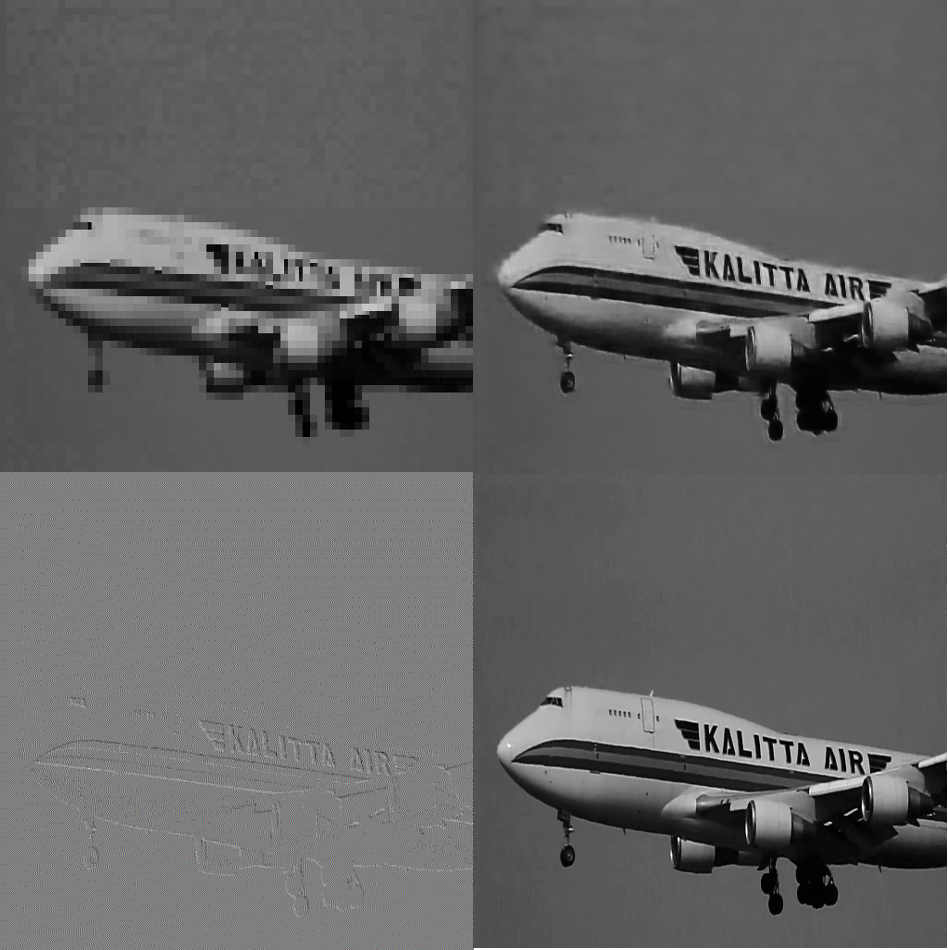}}
\caption{Reconstruction by edge enhancement algorithm \cite{chopp2021_events} for ILSVRC2015$\_$train$\_$01054000 sequence (frame $82$). Top left: Distorted frame, Bottom Left: Event frame, Bottom Right: Actual frame, Top Right: Edge enhanced reconstructed frame.}
\label{Event_rec_image}
\end{figure}

$S_t$, $Q_t$ and $V_t$ are sent from the chip to the host at time $t$. The reconstructed frame $\hat{f}_{t-1}$ at $t-1$ is used along with $S_t$, $Q_t$ and $V_t$ to reconstruct frame $\hat{f}_t$. Since the frame $\hat{f}_t$ is decomposed as QT blocks, it has block-like artifacts, clearly identifiable near the edges of the objects in the scene, especially at low bit rates, as shown in Fig. \ref{Event_rec_image} (top left). In order to enhance the quality of the edges in $\hat{f}_t$, we use an event-based edge enhancement algorithm.

\emph{Event-based edge enhancement:}
The events are used to enhance the edges in intensity frames using a deep neural network model, as detailed in \cite{chopp2021_events}. The last and current reconstructed frames $\hat{f}_{t-1}$ and $\hat{f}_t$ along with the event frames between $t-1$ and $t$ are fed as input to the model. In this work, $4$ event frames are fed to the model, which has residual blocks along with convolutional layers. The resulting enhanced frame $\hat{f}^{edge}_t$ at time $t$ has edges which closely resemble the actual ones. Figure \ref{Event_rec_image} shows an example of the edge enhancement for a highly distorted frame. The distorted frame $\hat{f}_{82}$ has significant distortion (generated with $\lambda = 650$), with the letters clearly not distinctly seen on the side of the airplane. With events, the resultant enhanced frame has edges which are quite distinct. The letters written on the body of the airplane are also clearly readable, with the block-like artifacts significantly reduced. By looking carefully at the edge enhanced frame $\hat{f}^{edge}_t$, it can be seen that the edges are preferentially enhanced especially on part of the image the events are concentrated at.

\emph{Object Detection and Tracking on Intensity Frames:} 
The regions of interest in the edge enhanced intensity frame $\hat{f}^{edge}_t$ are detected by using an object detector on the host. In this architecture the object detector is a modular sub-system. It can be updated/replaced/augmented as per the requirements of the end application. In this work Faster R-CNN is used as the object detector, with the detector trained with a novel 2-step methodology as in our previous work \cite{ReImagine_Sys_1}. The object detector generates bounding boxes with class labels, which are fed as input to an object tracker. Our modular design approach allows upgrading the object tracker as well. In this work, we use a Kalman filter-based multiple object tracker, simple online and realtime tracking \cite{SORT_2016} for its popularity and easy implementation, as used in our previous work \cite{ReImagine_Sys_1}. The key adaptation from \cite{SORT_2016} is that, in our implementation, the tracker appends the class information of the objects, which is critical for fusing the regions of interest as described in \ref{Fusion_RoI}. For every time $t$, the bounding boxes from the object detector are used as the observations for updating the state predicted by the Kalman tracker.  

\subsubsection{Event based object detection and tracking}
\label{Event-based Processing}
The compressed events are sent from the chip to the host for further processing. On chip, the events are sampled and temporally quantized to generate event frames. The events are received by the host as data packets, which are then decoded by inverting the coding processes of run length and Huffman encoding. For the intensity frames at $F_{i}$ frames per second (fps), the events are aggregated into event frames at $F_{e}$ fps. Thus, the events during $1/F_{e}$ seconds are aggregated together into event frames. Conversion of the events from asynchronous to synchronous event frames enables their frame-based processing.

\begin{figure*}[t]
\begin{multicols}{3}
    \noindent
    \includegraphics[width = \linewidth, height=3cm]{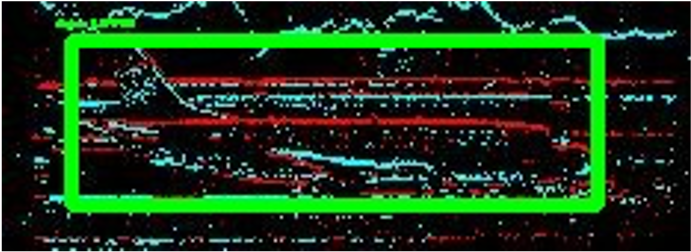}\par\caption*{(a) Airplane}
    \includegraphics[width = \linewidth, height=3cm]{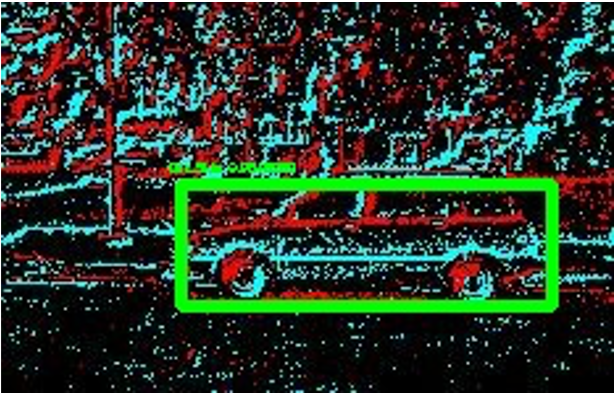}\par\caption*{(b) Car Side}
    \includegraphics[width = \linewidth, height=3cm]{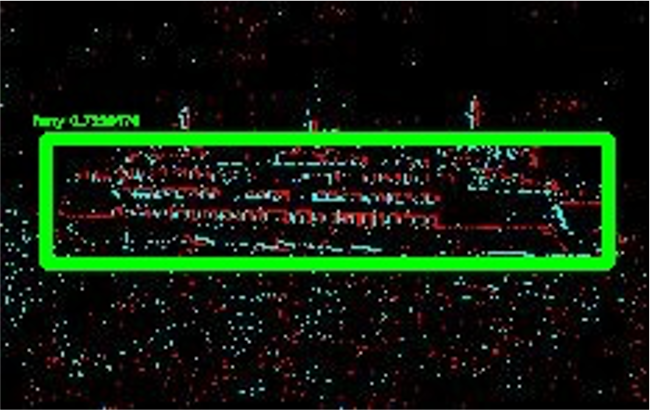}\par\caption*{(c) Ferry}
\end{multicols}
\caption{Examples of object detection in event frames from the N-Caltech dataset \cite{N-Caltech_2015}}
\label{fig_event_detections}
\end{figure*}

On host, the events can be used to perform any task specific processing. In this work, object tracking is the end task to be performed. Hence, the events are used to improve object tracking performance. For fast moving objects, tracking of objects using intensity frames only leads to mis-detection of objects owing to blur and fewer frames are able to capture the object moving across the field of view. The event frames are used to circumvent such situations, and accurately detect and identify objects using an event based object detector. The event based object detector helps in not only locating and identifying the fast moving objects which move significantly during $1/F_i$ seconds but also objects which disappear from the frame within $1/F_i$ seconds. This improves the tracking accuracy of the objects over the sequence of frames.  It must be kept in mind that any event based object detector can be used in our modular architecture for identifying objects from events. However, in this work, Tiny Yolo \cite{cannici_YOLE} has been used as the object detector on the aggregated event frames due to its light-weight architecture owing to fewer convolutional layers in the object detection network. It must be kept in mind that typically $F_{e} > F_{i}$. For instance, for $30$ fps intensity frame rate, the aggregated event frame rate could be $120$ fps (or higher) and hence the event based object detector needs to process many more event frames over the same duration. The requirement of having a light-weight event object detector architecture is hence vital. Fig. \ref{fig_event_detections} show examples of airplane, car side and ferry detections on N-Caltech dataset \cite{N-Caltech_2015} using this event object detector. It can be seen that the event detector is able to identify, detect and classify objects in both dense event  frames Fig. \ref{fig_event_detections} (b) (Car Side) as well as from relatively sparse event frames Fig. \ref{fig_event_detections}(a),(c).

The event object detector generates bounding boxes with class labels, which are used as input to the event object tracker. The class labels are appended for each of the bounding boxes tracked by the event tracker. We maintain a separate tracker for the objects detected from the event frames. This is due to the fact that the objects detected from intensity frames might be different from the objects detected from the event frames. Moreover the aggregated event frames are at a higher frame rate than the intensity frame rate. Thus the detection of objects from the event frames will be at a higher frame rate than intensity based detections. This will inherently introduce a mismatch in the rate of information received from each of the modalities to track objects in a scene.

\begin{figure}[htbp]
\centerline{\includegraphics[width=\linewidth]{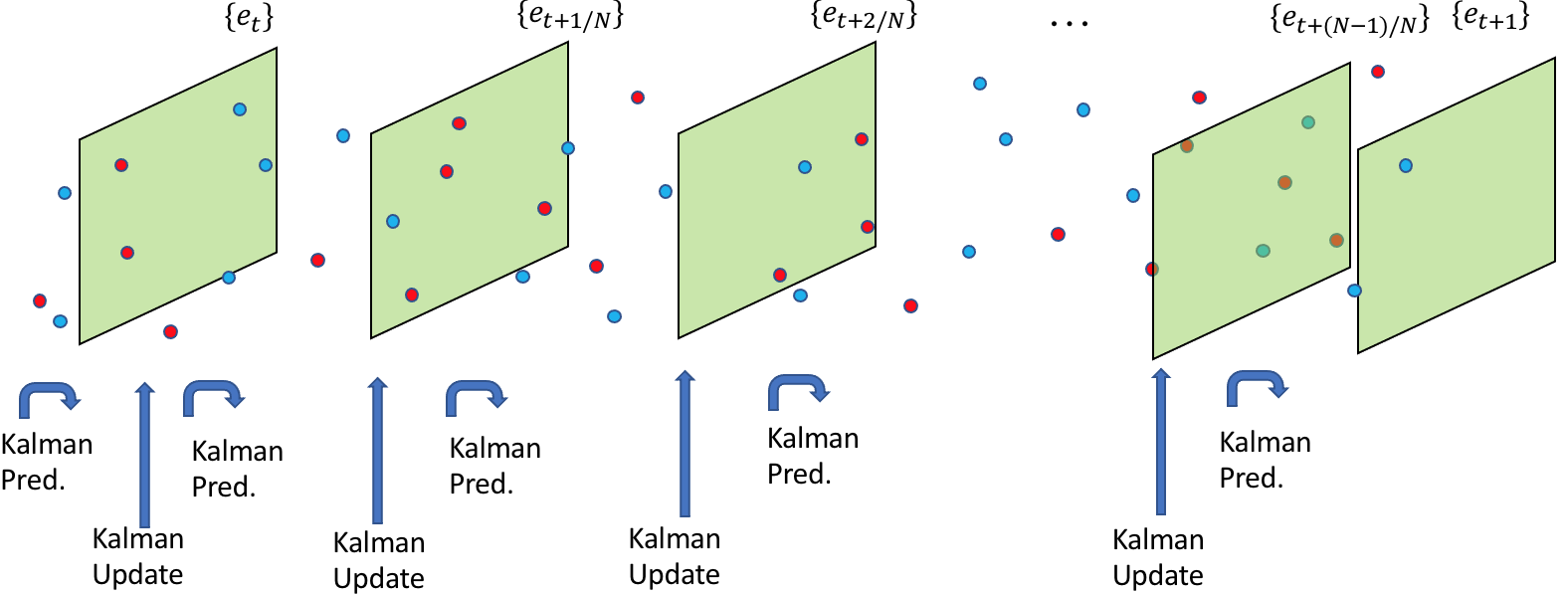}}
\caption{Event Kalman Filter tracking}
\label{Event_KF_tracking}
\end{figure}

The event based object tracker is a modular component of the whole architecture which can be replaced / updated based on the users' choice. In this work, we adapt the Kalman filter-based multiple object tracker, SORT \cite{SORT_2016} for object tracking similar to the implementation in intensity based object tracker in Section \ref{Intensity based Processing}. The event tracker uses a linear motion model to predict the bounding box location in the event frame ${\hat{e}}'_t$ based on the state at time $ t-1 + (N-1)/N$. The observations are the bounding boxes detected by the event object detector at time $t$ with the association of the observed and predicted bounding boxes done as described in Section \ref{Intensity based Processing}. The Kalman filter predicts the location of the bounding boxes, $\tilde{bb}^e_{t+1/N}$ at time $t + 1/N$. At time $t+1/N$, the observations (bounding boxes) are available from the event object detector to update the state of the event tracker. This step of predict-update is repeated for $N-1$ times in between time $t$ and $t+1$ before finally predicting the bounding boxes $\tilde{bb}^e_{t+1}$ at time $t+1$, as shown in Fig. \ref{Event_KF_tracking}.

\subsubsection{Fusion of Regions of Interest}
\label{Fusion_RoI}
The fusion of the predicted intensity and event modality bounding boxes at time $t+1$, $\tilde{bb}^i_{t+1}$ and $\tilde{bb}^e_{t+1}$, respectively, are very critical for determining the resulting predicted bounding boxes before sending them to the chip. This is vital for the information being sent to the chip for prioritization of bits in the RoIs at time $t+1$, in the Viterbi optimization algorithm. 

The fusion model consists of two parts, bounding box filter module and bounding box fusion module. In the bounding box filter module, the features are first extracted for each bounding box predictions. Next, a fully connected neural network is constructed to predict the confidence score for each bounding box prediction based on these extracted features, which measures the probability of each bounding box predictions belonging to a ground truth bounding box and removes the bounding box predictions with low confidence score. Then, the bounding box fusion module utilizes a designed fusion mechanism to fuse the remaining bounding box predictions. 

\begin{comment}
\begin{figure}[htbp]
\centerline{\includegraphics[width=\linewidth]{images/fusion_structure.jpg}}
\caption{Fusion model consisting of Bounding Box Filter and Bounding Box Fusion modules. Bounding box filter removes the bounding box predictions with low confidence score. Bounding box fusion module fuses the remaining bounding boxes together based on the designed fusion mechanism.}
\label{fusion_structure}
\end{figure}
\end{comment}

\emph{Bounding Box Filter Module:}
The Bounding Box Filter Module takes as input the bounding box predictions, $\tilde{bb}^i_{t+1}$ and $\tilde{bb}^e_{t+1}$. The inputs are at the bounding box level, where we only know information of the bounding box and source of the bounding box where it comes from (intensity based Kalman filter prediction or event based Kalman filter prediction). Firstly, the bounding box filter extracts key features of each bounding box predictions. We design seven key features of each bounding box predictions as described:

\emph{(a) Bounding Box Class}: The class of each input bounding box predictions, for example Airplane, Cars, and others. 

\emph{(b) Bounding Box Source}: The source represents where the bounding box prediction comes from. $0$ and $1$ represents the bounding box prediction from the intensity Kalman filter and event Kalman filter, respectively.

\emph{(c) Bounding Box Size}: The size of each input bounding box predictions. It is simply the area of each bounding box.

\emph{(d) Aspect Ratio (AR)}: The AR of each class (example, Airplane) usually differs from another. AR, defined as ratio of height to width is chosen as a feature.

\emph{(e) Overlap Ratio (OR)}: OR is defined as the maximum Intersection over Union (IoU) between the target bounding box prediction and other bounding box predictions from the same source (intensity or events) as shown in Eqn. \ref{eq:overlap_ratio}. High OR indicates that the probability of those two bounding box predictions containing the same object is high.
\begin{equation}
    Overlap~Ratio = \max \limits_{j \neq i}{IoU(BBP(i),BBP(j))}
    \label{eq:overlap_ratio}
\end{equation}

\emph{(f) Crowdedness Ratio (CR)}: The CR measures the crowdedness of each bounding box predictions from the same source. It is the number of the other bounding box prediction centers from the same source in the target bounding box regions.

\emph{(g) Support Value (SV)}: For a given bounding box from a particular source, if there exists a bounding box prediction from the other source and its IoU with the target bounding box prediction is greater or equal to threshold ($0.7$), the SV for the target bounding box prediction is $1$. Otherwise, there is no bounding box prediction from the other source, and the SV is $0$. If SV is greater than $0$, we can find a correlated bounding box prediction from the other source, which means the intensity Kalman filter and event Kalman filter predictions detect the same object, and the probability of such bounding box prediction is relatively high. 

After obtaining the feature vector for each predicted bounding box using the feature extractor, we construct a neural network with three fully connected layers to predict the confidence score of each bounding box predictions, which measures the probability of each bounding box predictions belonging to a ground truth. We utilize a predefined threshold (of $0.7$) to filter out the bounding box predictions with low confidence score. 

\begin{figure}[h]
\begin{center}
\includegraphics[width=\linewidth]{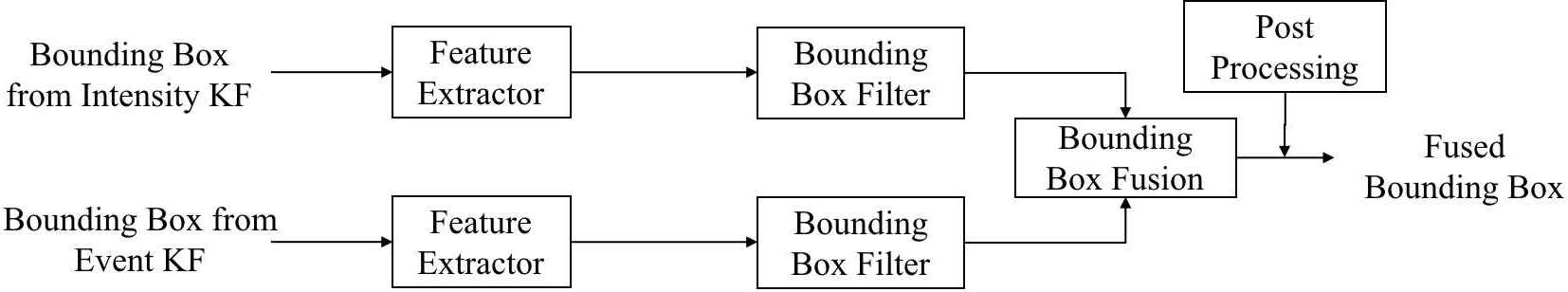}
%\vspace*{-5mm}
\end{center}
  \caption{Workflow of the Fusion Model. (KF: Kalman Filter)}
\label{bounding_box_filter}
%\vspace*{-5mm}
\end{figure}

\emph{Bounding Box Fusion Module}
The filtered bounding box predictions is aligned with the non-filtered intensity Kalman filter and event Kalman filter predictions, and the maximum IoU value is computed. If the maximum IoU is greater than or equal to the threshold of $0.5$, both Kalman filter predictions are considered to detect this filtered bounding box object, and, we have the option of utilizing three fusion strategies to fuse the Bounding Box predictions together: Intersection (bounding box as the intersection region), Union (bounding box as the union region), and, Confidence (bounding box as the one with highest confidence), as shown below in Fig. \ref{fusion}. Otherwise, if the maximum IoU value is less than $0.5$, the filtered bounding box prediction is retained. 
\begin{comment}
\begin{itemize}
    \item \textbf{Intersection} : The intersection region is considered as the fused bounding box prediction.
    \item \textbf{Union} : The union region is considered as the fused bounding box prediction.
    \item \textbf{Confidence} : The bounding box with highest confidence score is considered as the fused bounding box prediction.
\end{itemize}
\end{comment}
\begin{figure}[h]
\begin{center}
\includegraphics[width=\linewidth]{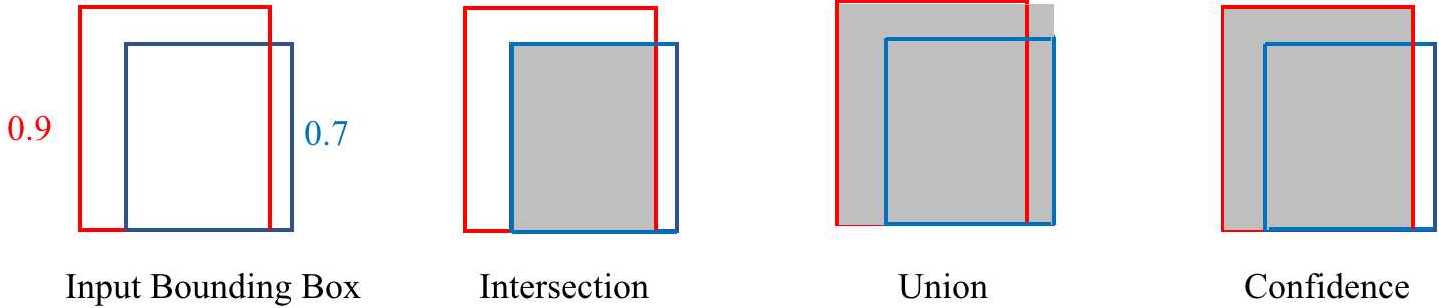}
%\vspace*{-5mm}
\end{center}
  \caption{Three Fusion Strategies. Red: Bounding Box from Intensity KF, Blue: Bounding Box from Event KF, Grey: Fused Bounding Box. (KF: Kalman Filter)}
\label{fusion}
%\vspace*{-5mm}
\end{figure}
Thus, we obtain the fused Bounding Box predictions. However, here we make use of the spatial information only. In order to exploit additional temporal information, we develop a post processing algorithm to filter out the false positive predictions. We assume the object movements during the successive frames to be relatively small, so the fused bounding box predictions in frame $t$ is assumed to be correlated with the fused bounding box predictions in the previous frame $t-1$. We compare the IoU value with all the fused Bounding Box predictions at frame $t$ with those at frame $t-1$. If the IoU is $\geq 0.5$, the fused Bounding Box prediction at frame $t$ follows the temporal correlation and we keep this Bounding Box prediction as the output. Otherwise, the fused bounding box predictions are a false positives, and are filter out. The workflow of the whole Fusion Model is shown in Fig. \ref{bounding_box_filter}. 
\subsection{Synchronization and Timing Diagram}
\begin{figure}[htbp]
\centerline{\includegraphics [width=0.85\linewidth]{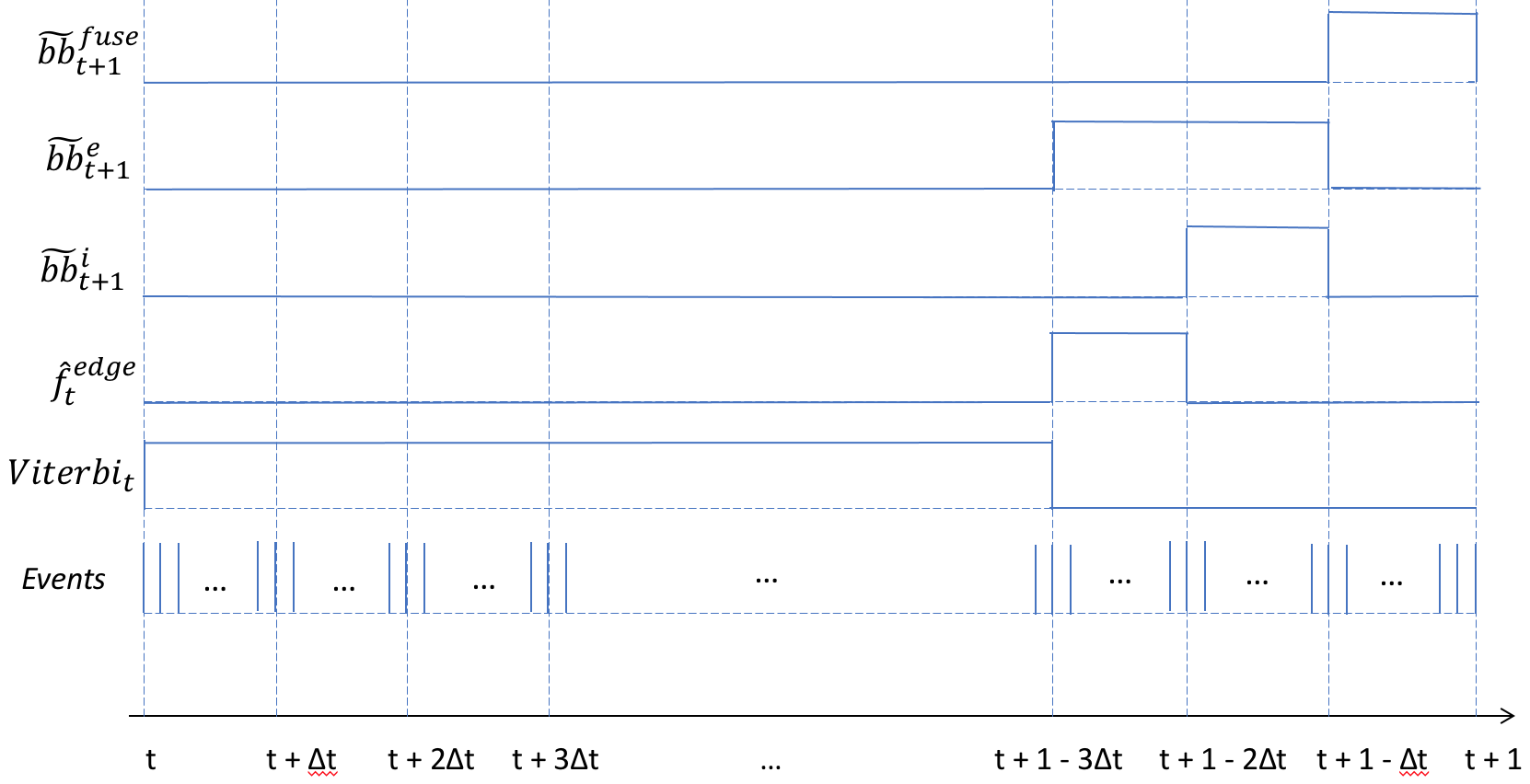}}
\caption{Timing Diagram for the Host-Chip Architecture}
\label{fig_time_sync}
\end{figure}

In this proposed system, the time synchronization of the different sub-processes need to be completed sequentially before the subsequent process can begin. The timing diagram is shown in Fig. \ref{fig_time_sync}. For every intensity frame at times $t$ and $t+1$, the events are acquired on the chip asynchronously. However, the Viterbi optimization algorithm starting at time $t$ must be synchronized in such a way that the optimization can be performed on the frame acquired at time $t$, alongwith the events acquired on the chip between time $t-1$ and $t$. The Viterbi optimization ends at time $t+1-3\Delta t$ and data is transferred from chip to host. The edge reconstruction computation on the host starts once the host receives the compressed intensity and event data, and generates enhanced frame $\hat{f}^{edge}_t$, at time $t+1-2\Delta t$. The intensity based object detector operates on the enhanced intensity images and passes the bounding boxes to the intensity based Kalman Filter tracker as its observations. The intensity based Kalman Filter generates predicted bounding boxes for time $t+1$, $\tilde{bb}^i_{t+1}$ at time $t+1-\Delta t$. On the other hand, the event based object detector computes the bounding boxes and updates the event based Kalman filter tracker $F_{e}/F_{i}$ times. This computation needs to be completed before time $t+1 - \Delta t$ such that the fusion model can take the predicted intensity and event bounding boxes at time $t+1 - \Delta t$ and finish computation at $t+1$ such that the predicted bounding box $\tilde{bb}^{fuse}_{t+1}$ is available for the Viterbi optimization at $t+1$ on the chip.
It must be highlighted that for illustration purposes we have neglected the time required for transmission of data between the chip to the host and vice-versa. In practice, depending on the network congestion, some delay may be introduced which can lead to minor changes in the timing diagram of Fig. \ref{fig_time_sync}.

\subsection{Peformance Evaluation Metric}
The performance of the complete host-chip tracking system on resource-constrained chip device has been evaluated in terms of the Multiple Object Tracking Accuracy (MOTA) evaluation metric \cite{bernardin2006multiple}. The MOTA metric is defined as 
\begin{align}
\label{eqn:eqlabel15}
MOTA = 1 - \sum_{t} {\dfrac{m_t + fp_t + mme_t}{g_t}},
\end{align}
where $m_t$ represents the number of missed detections,
${fp}_t$ the number of false positives,
$mme_t$ the number of mismatch (track switching) errors, and,
$g_t$ the number of ground truth objects, all at time $t$.
A high $MOTA$ score implies a high tracking accuracy of the objects and vice-versa.

\section{Experimental Results}
\subsection{Dataset}
\label{dataset_gen}
The proposed host-chip system requires intensity frames and asynchronous events from the same scene during evaluation. In order to do so, our framework requires intensity frames and events from the same dataset for training, validation and testing various parts of the system. The lack of large dataset for intensity and events is a key issue in literature, and we use abundantly available intensity datasets to generate events. We generate event data using Super SloMo algorithm \cite{jiang2018super} and ESIM \cite{rebecq2018esim} as in \cite{gehrig2020video}. Our algorithm can work with any object detection and tracking video datasets with one or multiple classes of tracking objects. However, in this work, for illustration purposes, we use the Airplane, Watercraft and Car classes from the ImageNet (ILSVRC) VID dataset \cite{russakovsky2015imagenet} to demonstrate results of our algorithm.
\begin{figure}[htbp]
\centerline{\includegraphics [width=\linewidth]{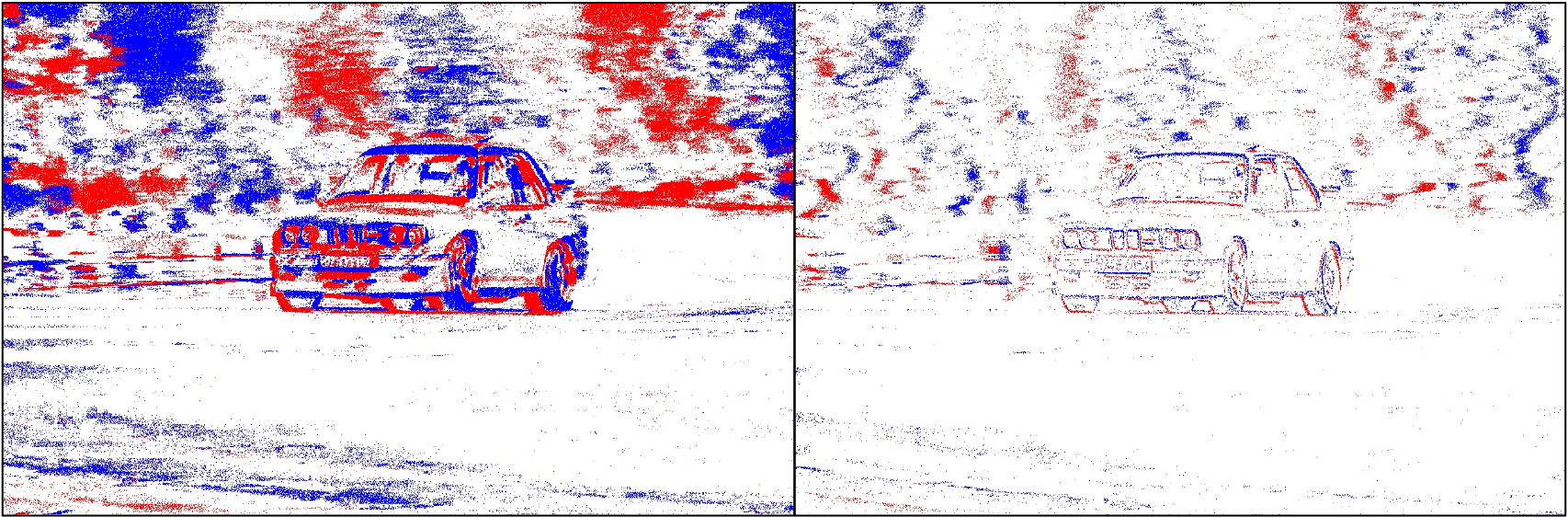}}
\caption{Sample Event Frame without (left) and with (right) frame interpolation. Blue:positive events, Red:negative events.}
\label{Event_frame_wo_w_intp}
\end{figure}
\begin{figure*}[t]
\begin{multicols}{3}
    \noindent
    \includegraphics[width = \linewidth, height=4.5cm]{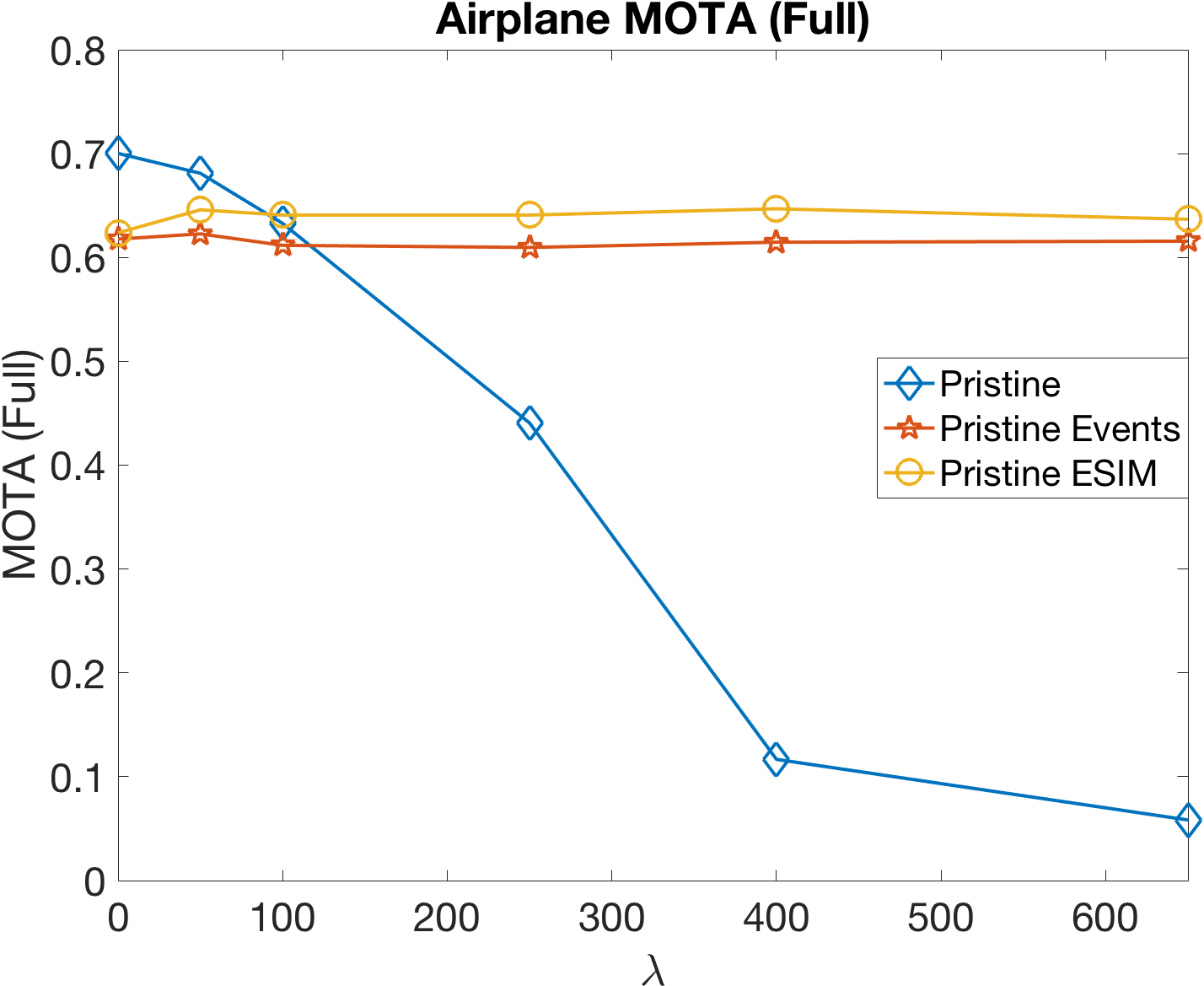}\par\caption*{(a) Airplane}
    \includegraphics[width = \linewidth, height=4.5cm]{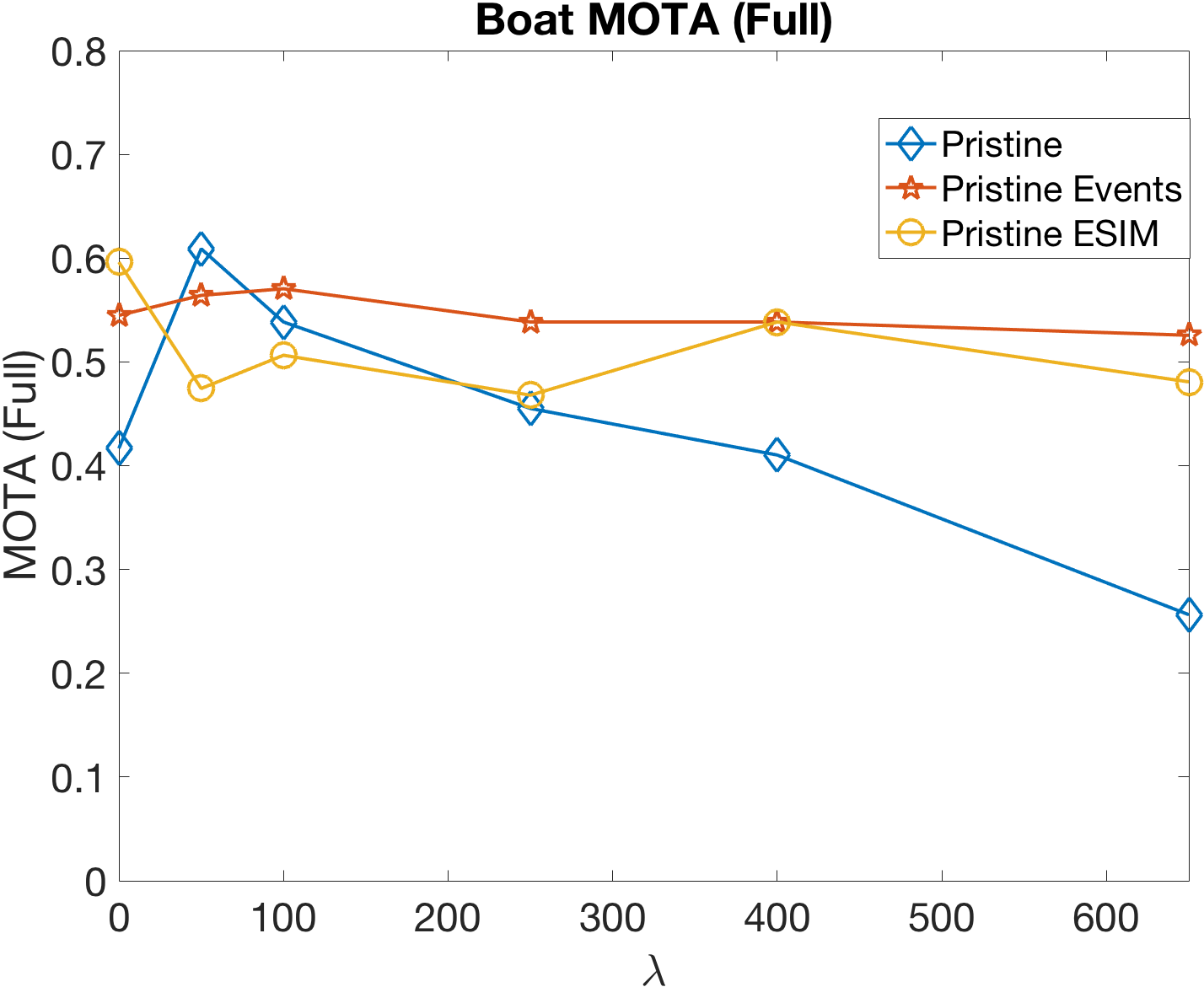}\par\caption*{(b) Watercraft}
    \includegraphics[width = \linewidth, height=4.5cm]{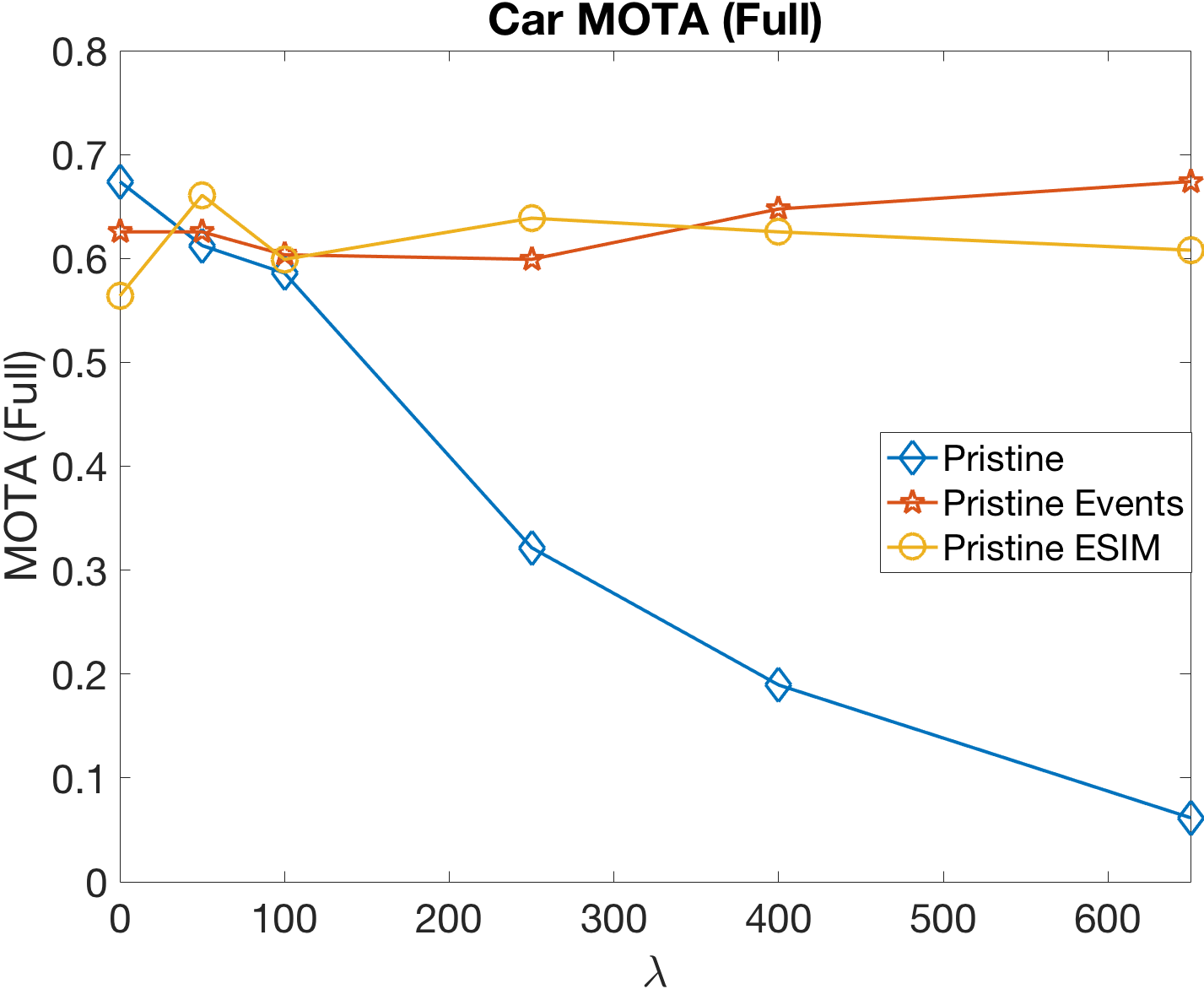}\par\caption*{(c) Car}
\end{multicols}
\caption{MOTA score of sequences with and without edge enhancement}
\label{fig_edge_recon}
\end{figure*}

A two-step approach is followed. Firstly, we interpolate the intensity to 240 fps using Super SloMo algorithm \cite{jiang2018super}. We use Super SloMo algorithm \cite{jiang2018super}, since this popular algorithm allows intensity frame interpolation to any arbitrary frame rate. We choose an upsampling factor of around $8\times$ as not being too high or too low. Too high of a factor may be a computational burden while too low of a factor might cause aliasing of the intensity signal \cite{gehrig2020video}. Additionally, we also interpolate the bounding box annotations along with the class labels to 240 fps from the bounding boxes at the frame rate in the dataset (typically $25 /30$ fps). Secondly, we use event simulation ESIM \cite{rebecq2018esim} for generating events from the interpolated intensity frames, keeping positive and negative thresholds at 0.15. In this two-step process, interpolating of intensity frames is critical, as the event frames generated without interpolated frames tend to be quite "thick" and appears visually as artificial edges, while the event frames generated with frame interpolation are quite "thin", and appears visually to be close to the actual edges. Such a sample of an event frame is shown in Fig. \ref{Event_frame_wo_w_intp}. We show the experimental performance of the system by testing the model on these ILSVRC VID sequences: (i) a video of airplanes, ILSVRC2015$\_$val$\_$00007010.mp4; (ii) a video of watercraft, ILSVRC2015$\_$val$\_$00020006.mp4; and, (iii) a video of cars, ILSVRC2015$\_$val$\_$00144000.mp4, which we refer to as Airplane, Boat and Car sequences, respectively. The uncompressed bit rate considering the intensity frames only is $62.91$ Mbps. We show the system performance by performing experiments at $1$ Mbps and $1.5$ Mbps, which results in compression ratios of 62.91 and 41.94, respectively. 
\begin{figure*}[t]
\begin{multicols}{3}
    \noindent
    \includegraphics[width = \linewidth, height=4.5cm]{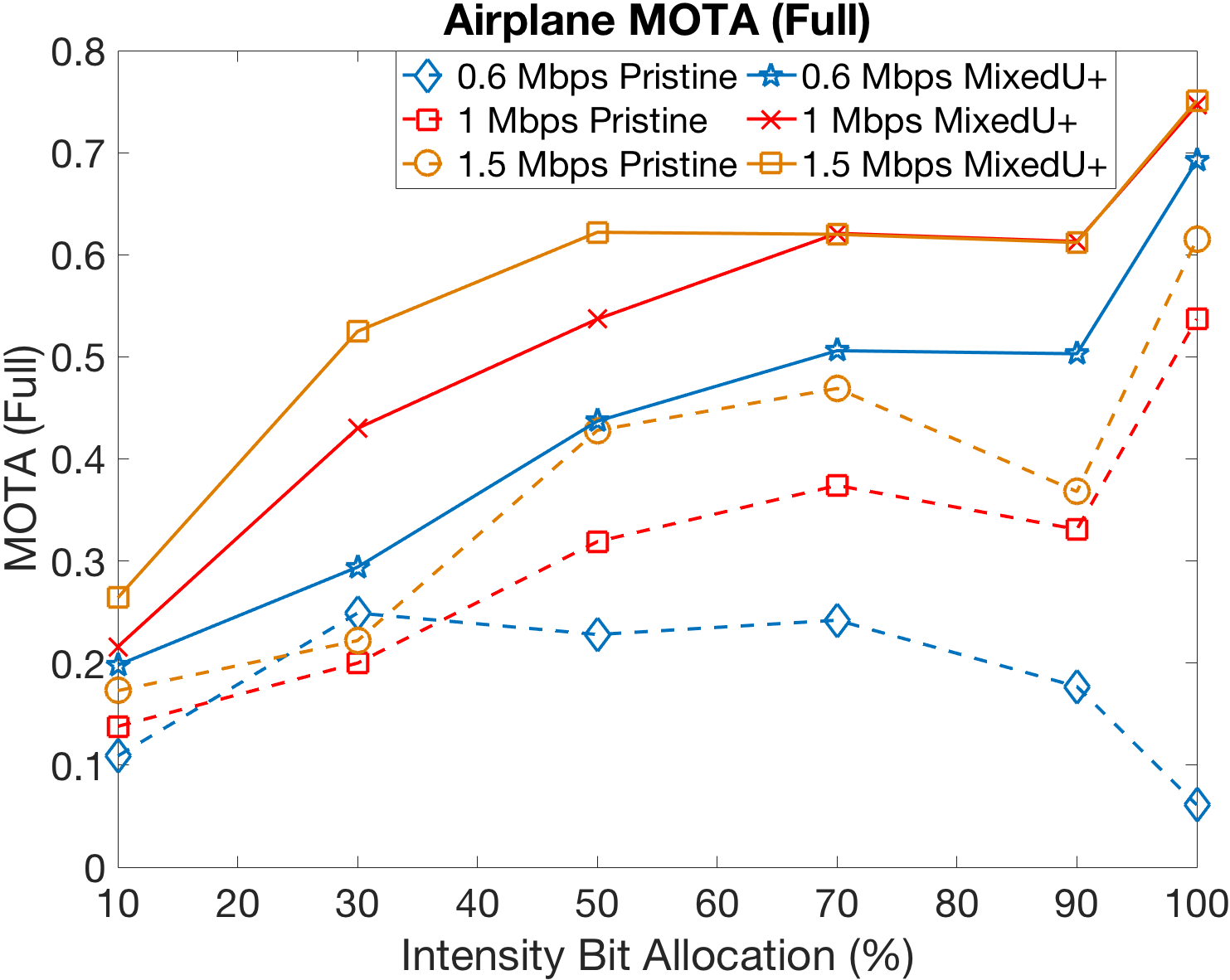}\par\caption*{(a) Airplane}
    \includegraphics[width = \linewidth, height=4.5cm]{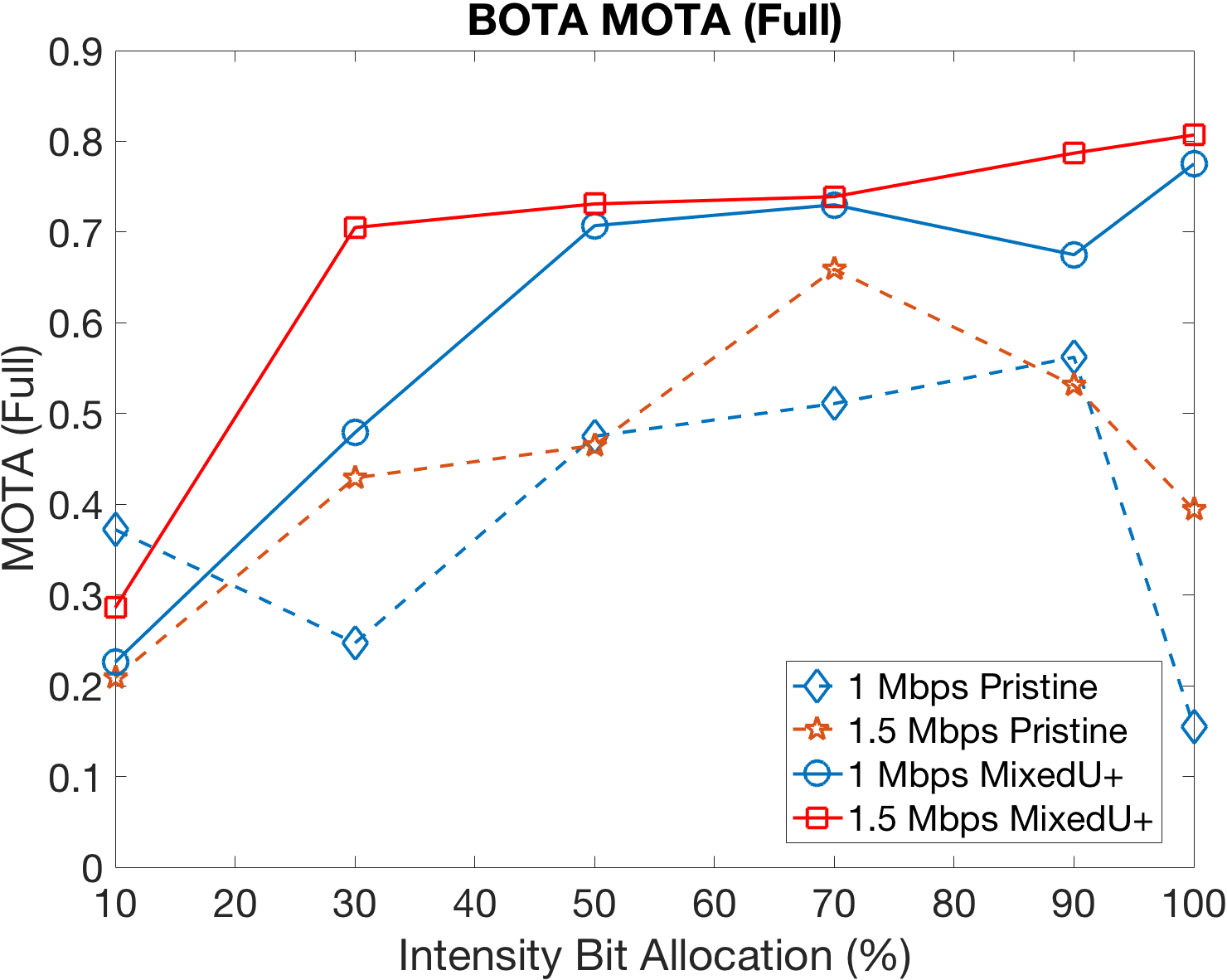}\par\caption*{(b) Watercraft}
    \includegraphics[width = \linewidth, height=4.5cm]{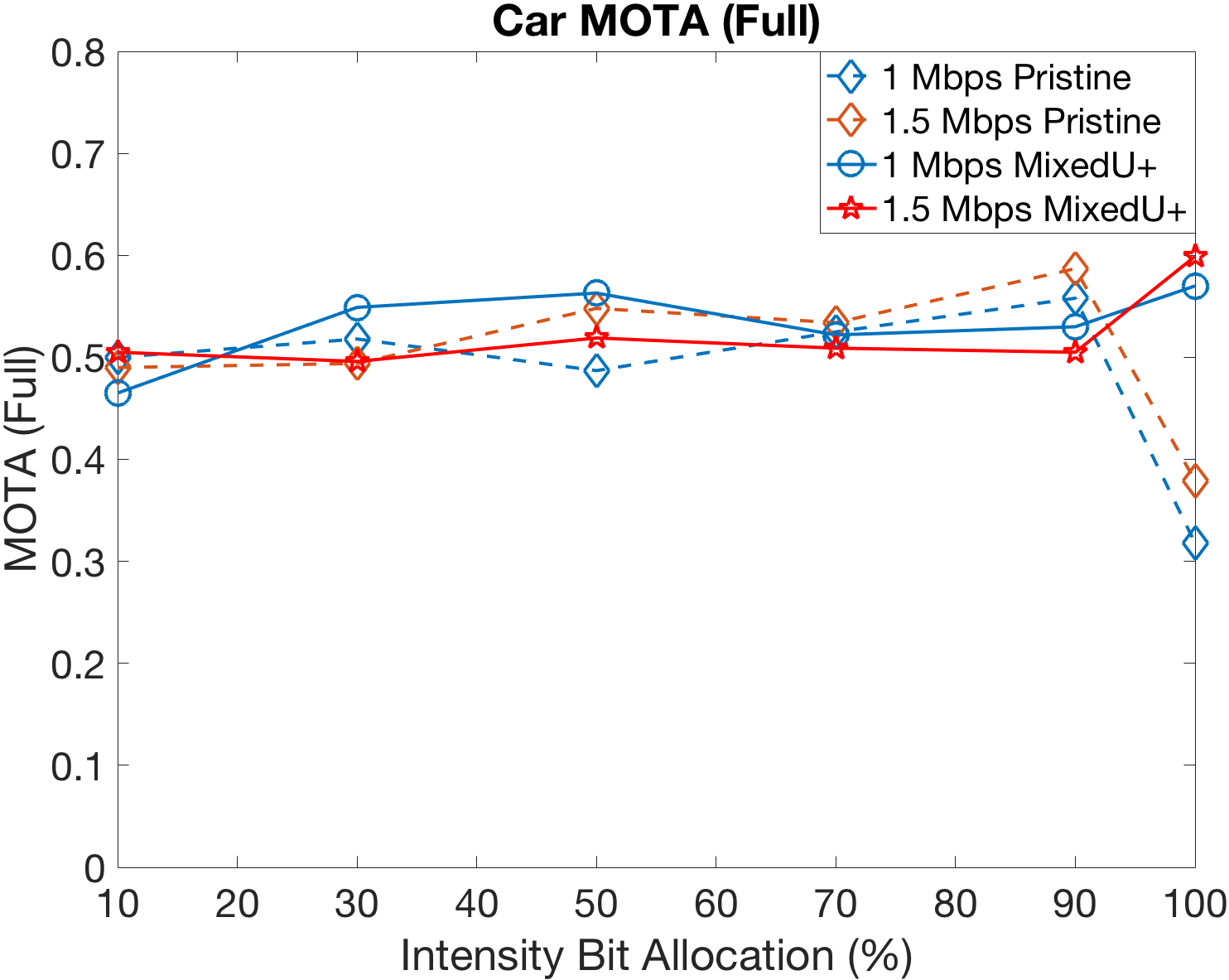}\par\caption*{(c) Car}
\end{multicols}
\caption{MOTA$_{full}$ of sequences with MixedU+ and Pristine detectors \cite{ReImagine_Sys_1}.}
\label{fig_edge_rec_bit_Rate}
\end{figure*}
\subsection{Performance with and without Edge Enhancement only}
\label{Performance_edge_enhancement}
In order to evaluate the contribution of the edge enhancement network in the overall system performance, we use the host-chip system with only edge enhancement without the use of events for object detection and tracking. We compare the system performance without edge enhancement. Original events have been used in these experiments with no distortion. The system performance metric (MOTA) is shown in Fig. \ref{fig_edge_recon}. The pristine object detector trained with undistorted frames of ILSVRC VID dataset, as described in \cite{ReImagine_Sys_1} is used as the base detector and edge enhanced intensity frames are detected using this pristine detector. Fig. \ref{fig_edge_recon} shows the system performance MOTA as a function of $\lambda$. A higher value of $\lambda$ in Equation \ref{eqn:eqlabel4} implies higher distortion and lower bit rate. It is seen that for the pristine object detector with no edge enhancement, MOTA reduces as $\lambda$ increases. We compare the edge enhancement results with the edge enhancement network trained by slightly two different event data: (a) event frames generated with simple intensity frame difference, and, (b) edge enhancement network trained on event frames generated using frame interpolation and ESIM successively as described in Section \ref{dataset_gen} (referred to as Pristine Events and Pristine ESIM, respectively, in Fig. \ref{fig_edge_recon}). The system is tested with actual events.
\begin{figure*}[t]
\begin{multicols}{3}
    \noindent
    \includegraphics[width = \linewidth, height=4.5cm]{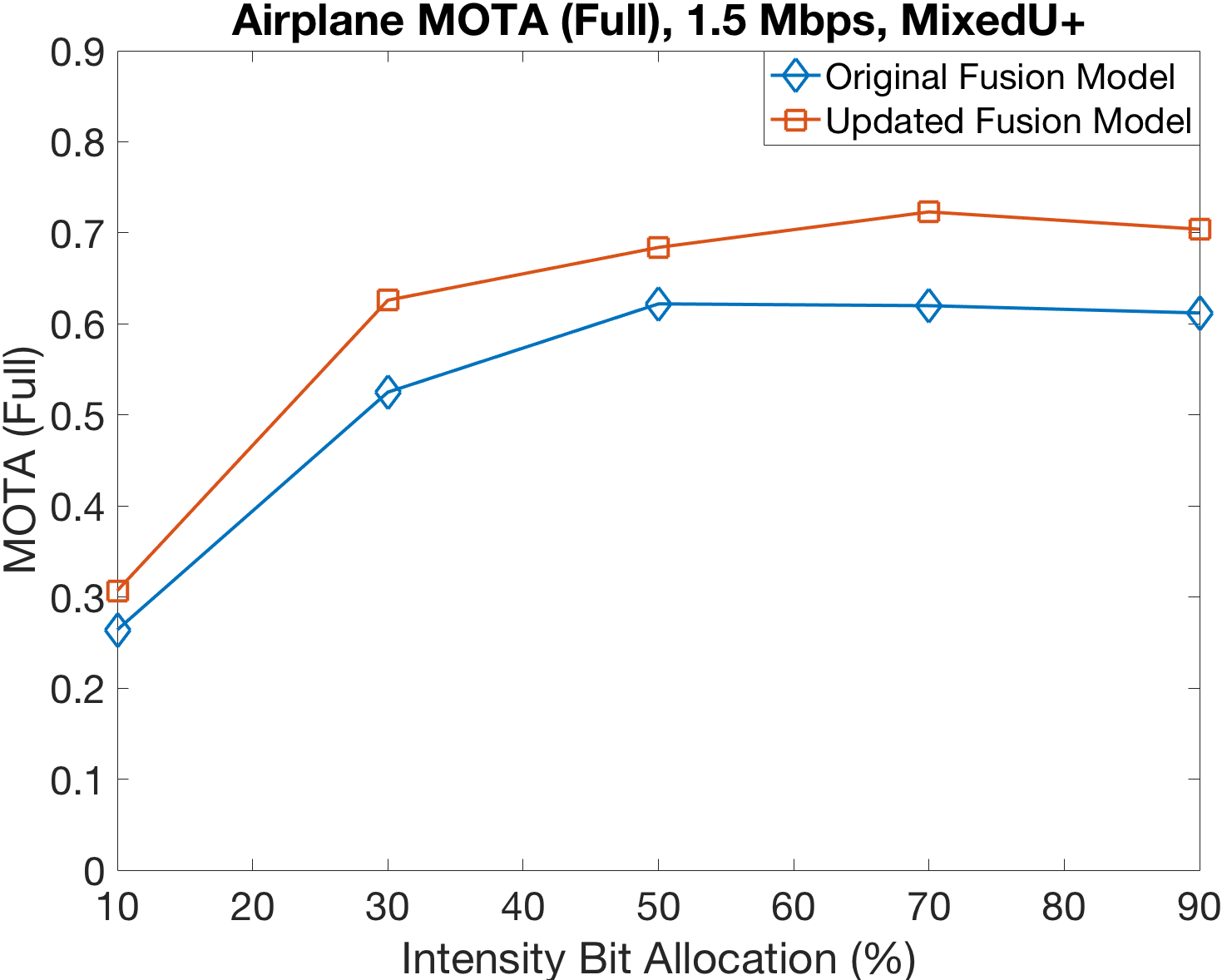}\par\caption*{(a) Airplane}
    \includegraphics[width = \linewidth, height=4.5cm]{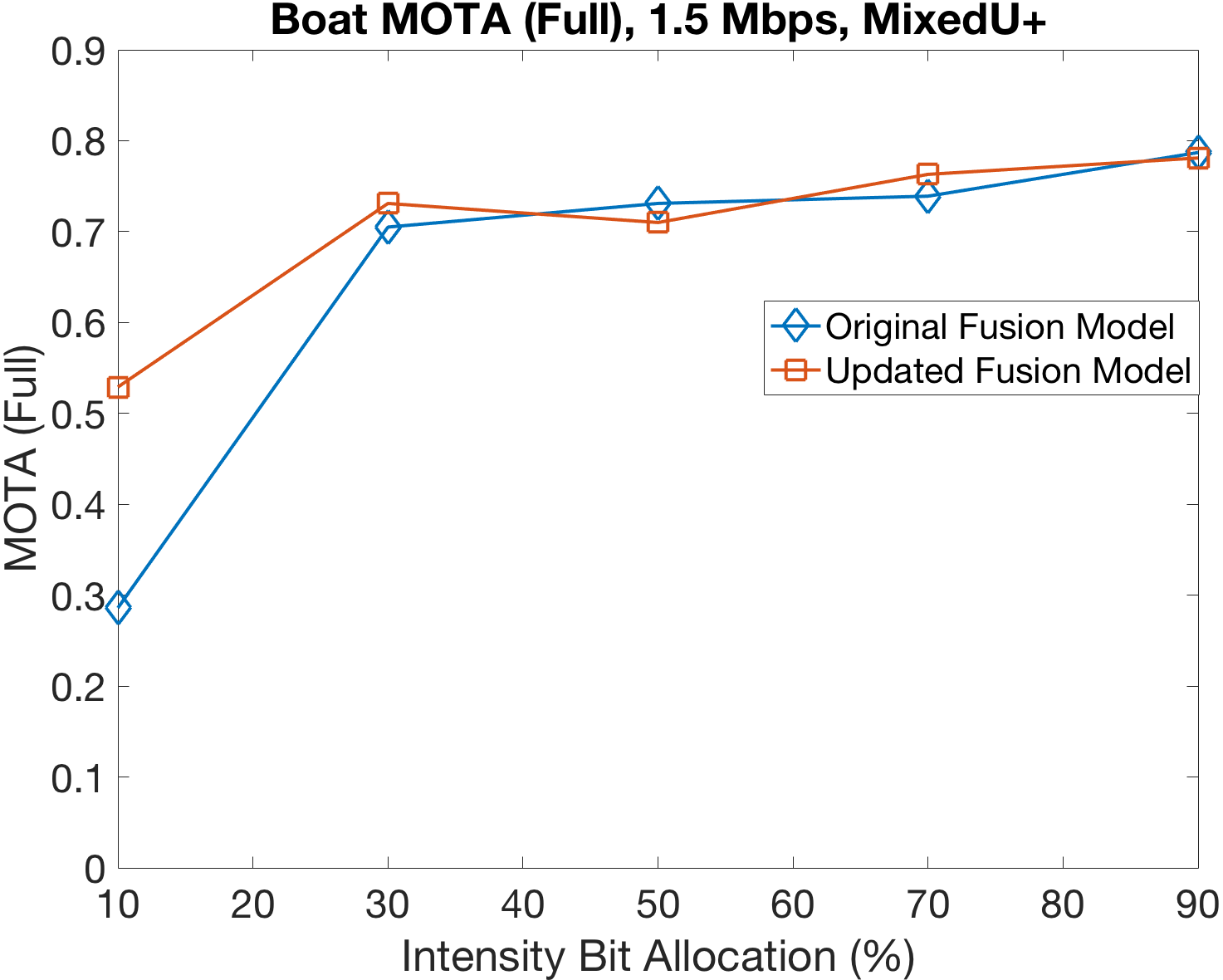}\par\caption*{(b) Watercraft}
    \includegraphics[width = \linewidth, height=4.5cm]{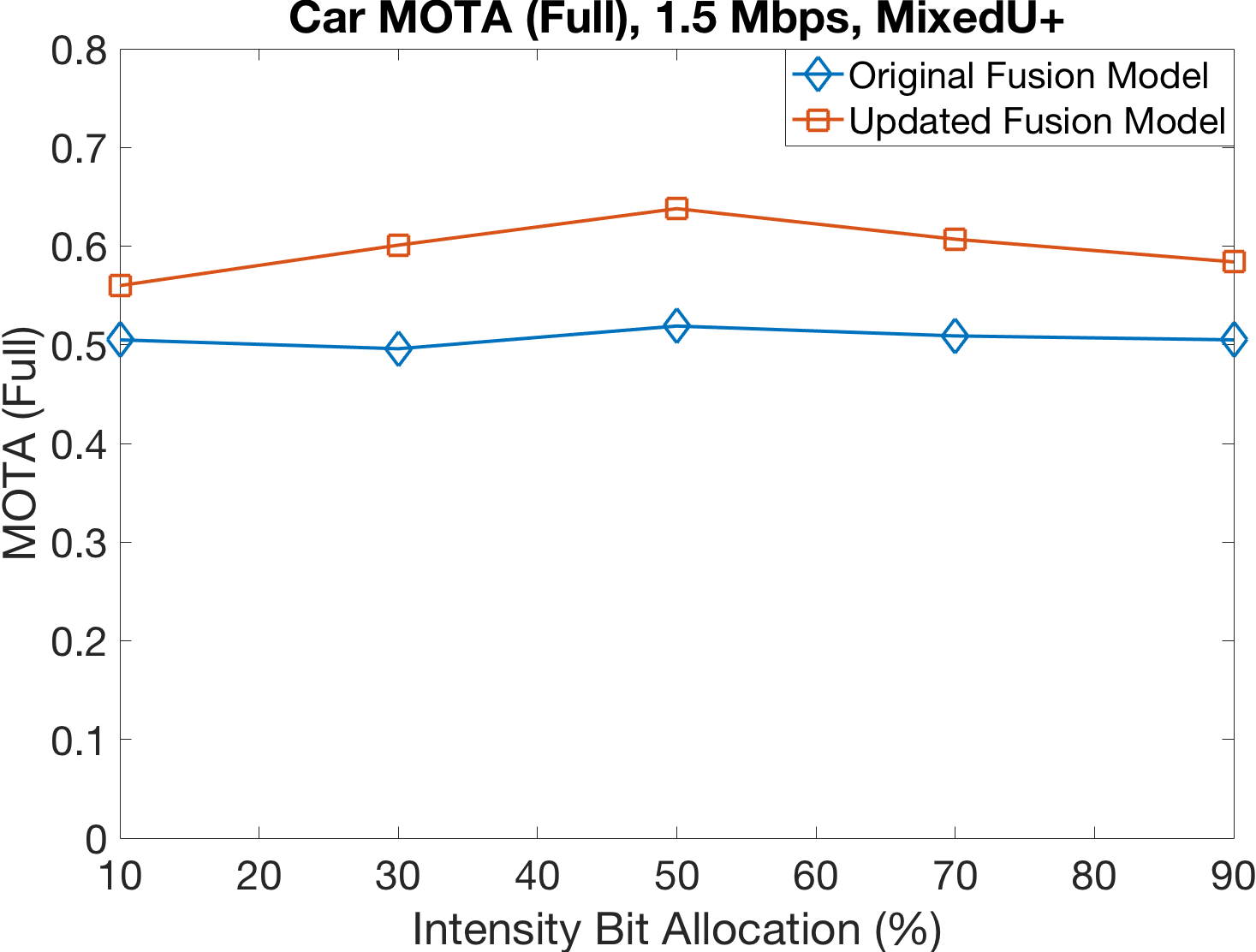}\par\caption*{(c) Car}
\end{multicols}
\caption{MOTA$_{full}$ of sequences with and without fusion network. Original Fusion Model and Updated Fusion Model refers to the system with and without event object detector, tracker and fusion network respectively.}
\label{fig_fusion}
\end{figure*}
\begin{figure*}[t]
\begin{multicols}{3}
    \noindent
    \includegraphics[width = \linewidth, height=4.5cm]{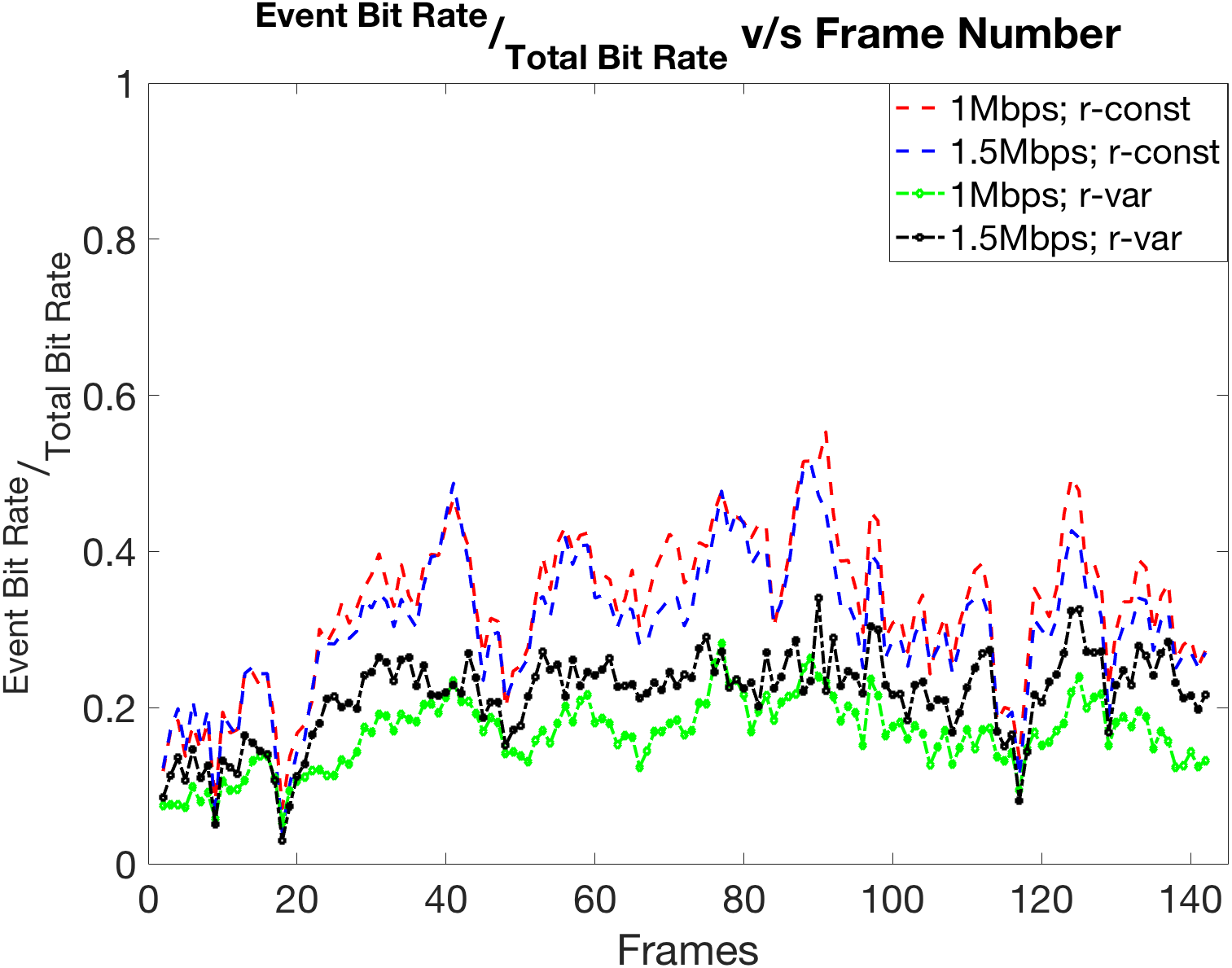}\par\caption*{(a) Airplane}
    \includegraphics[width = \linewidth, height=4.5cm]{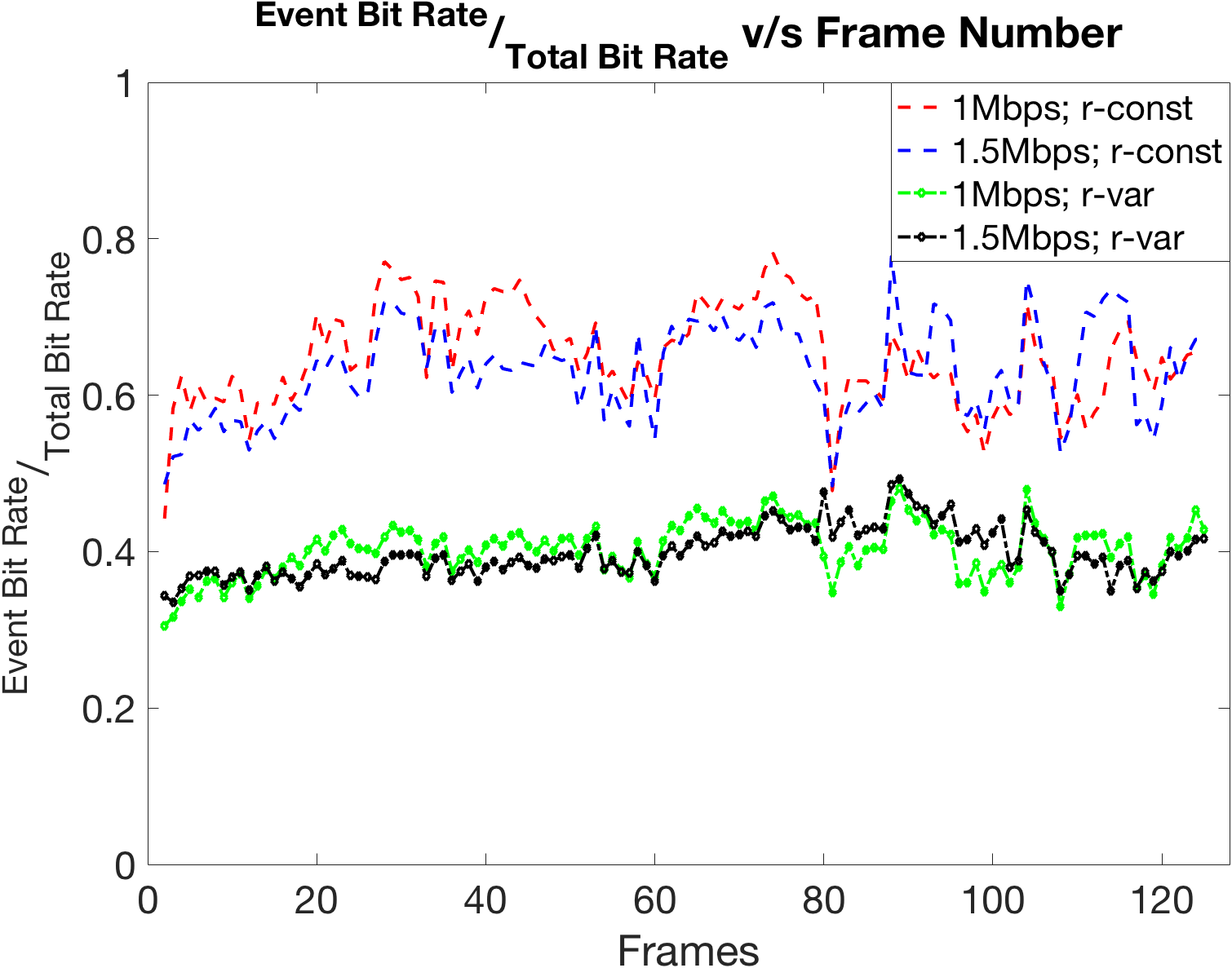}\par\caption*{(b) Watercraft}
    \includegraphics[width = \linewidth, height=4.5cm]{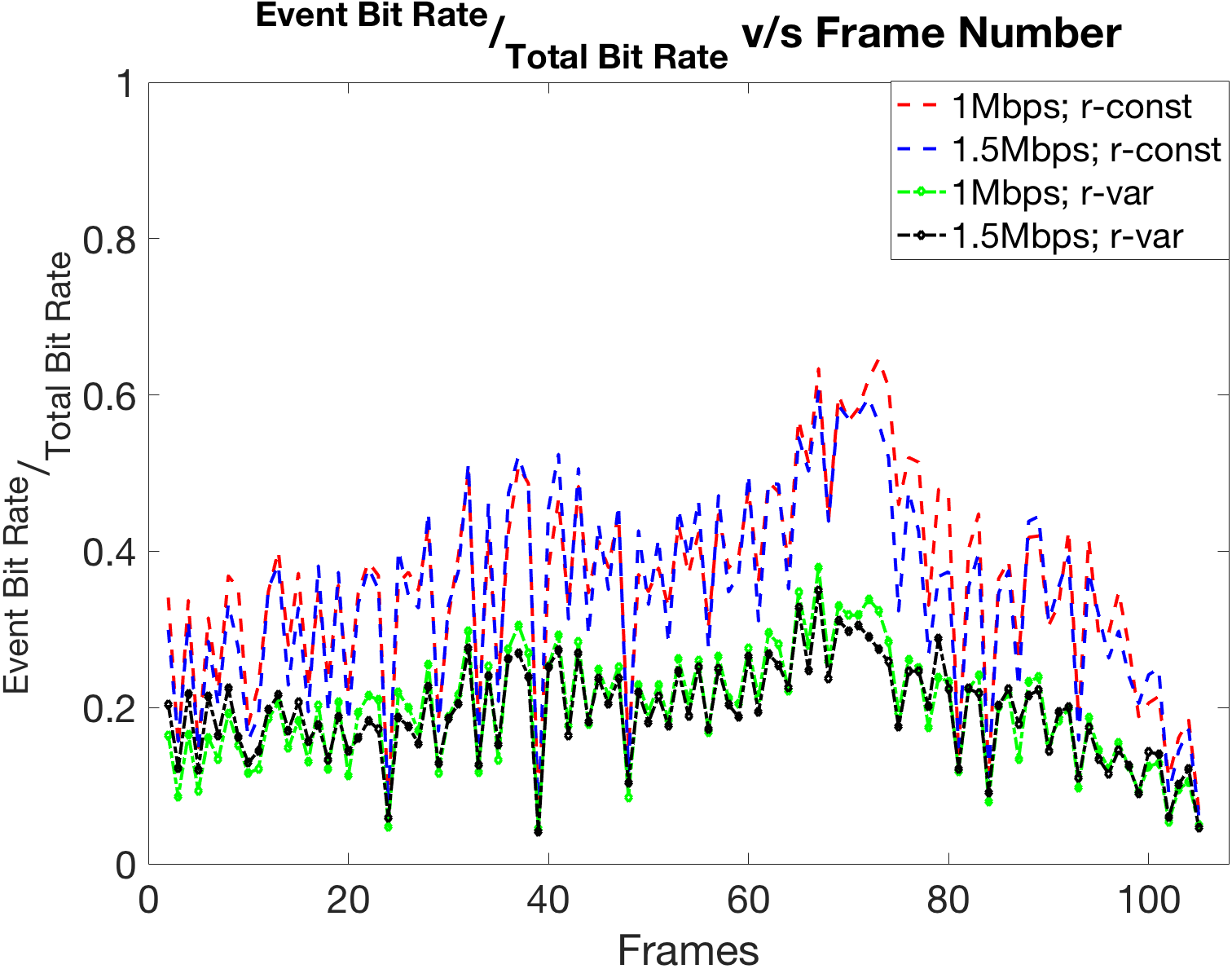}\par\caption*{(c) Car}
\end{multicols}
\caption{Comparison of allocated fraction of bits for constant and variable $r$.}
\label{fig_bit_rate_comp}
\end{figure*}
Clearly, the edge enhancement improves system performance. Additionally, it is seen that the performance of the system with Pristine ESIM is better than Pristine Events for the airplane and car sequences than with the boat sequence. However, the MOTA metrics using Pristine Events and Pristine ESIM are comparable, with Pristine ESIM performing at least or as good as Pristine Events for $61 \%$ of the cases in the airplane, boat and car sequences. In the subsequent experiments, unless mentioned otherwise, we use the ESIM version of edge reconstruction network due to its better performance, in addition to the fact that the training and testing with event data are identical for ESIM edge reconstruction network. 
\begin{figure*}[t]
\begin{multicols}{3}
    \noindent
    \includegraphics[width = \linewidth, height=4.5cm]{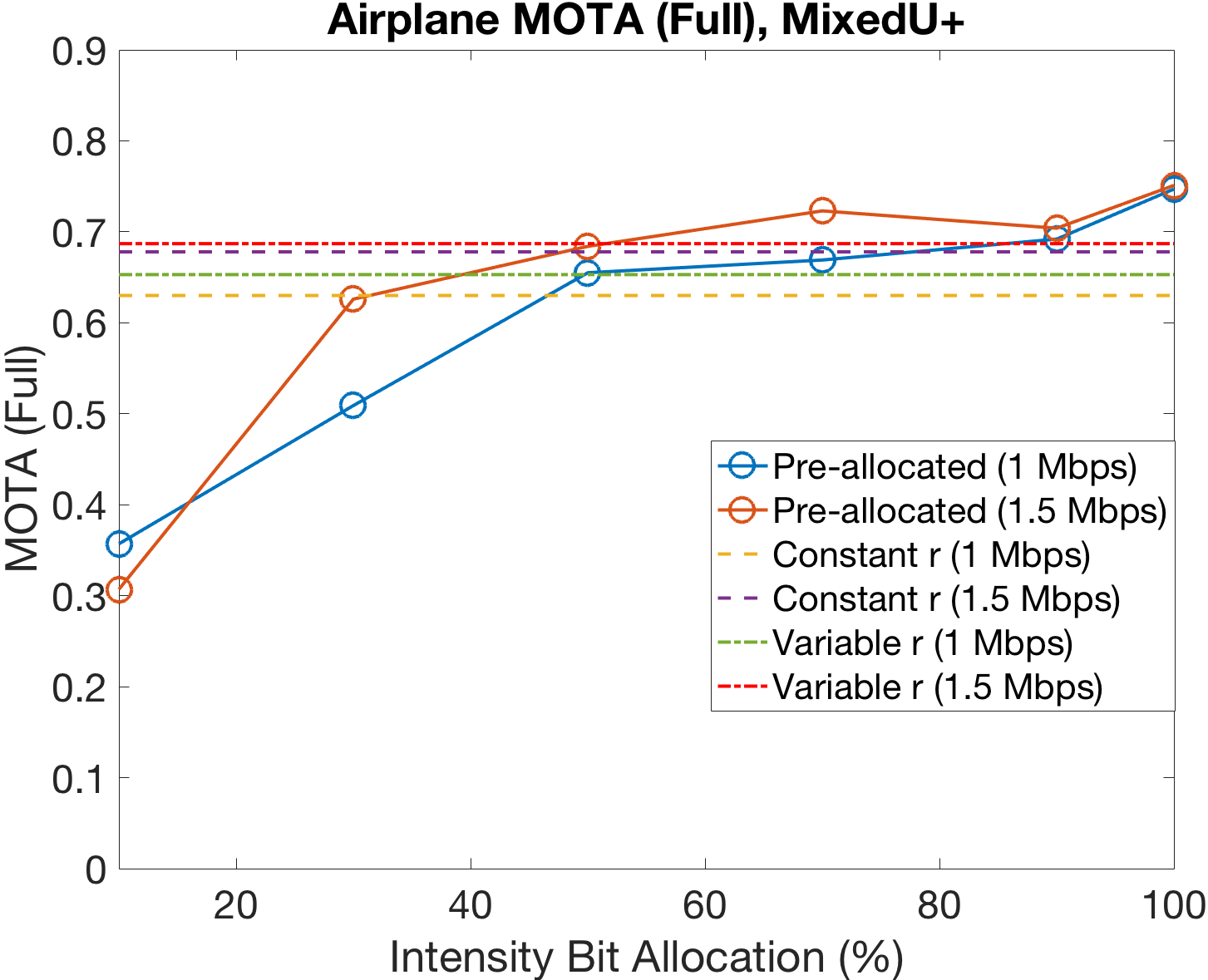}\par\caption*{(a) Airplane}
    \includegraphics[width = \linewidth, height=4.5cm]{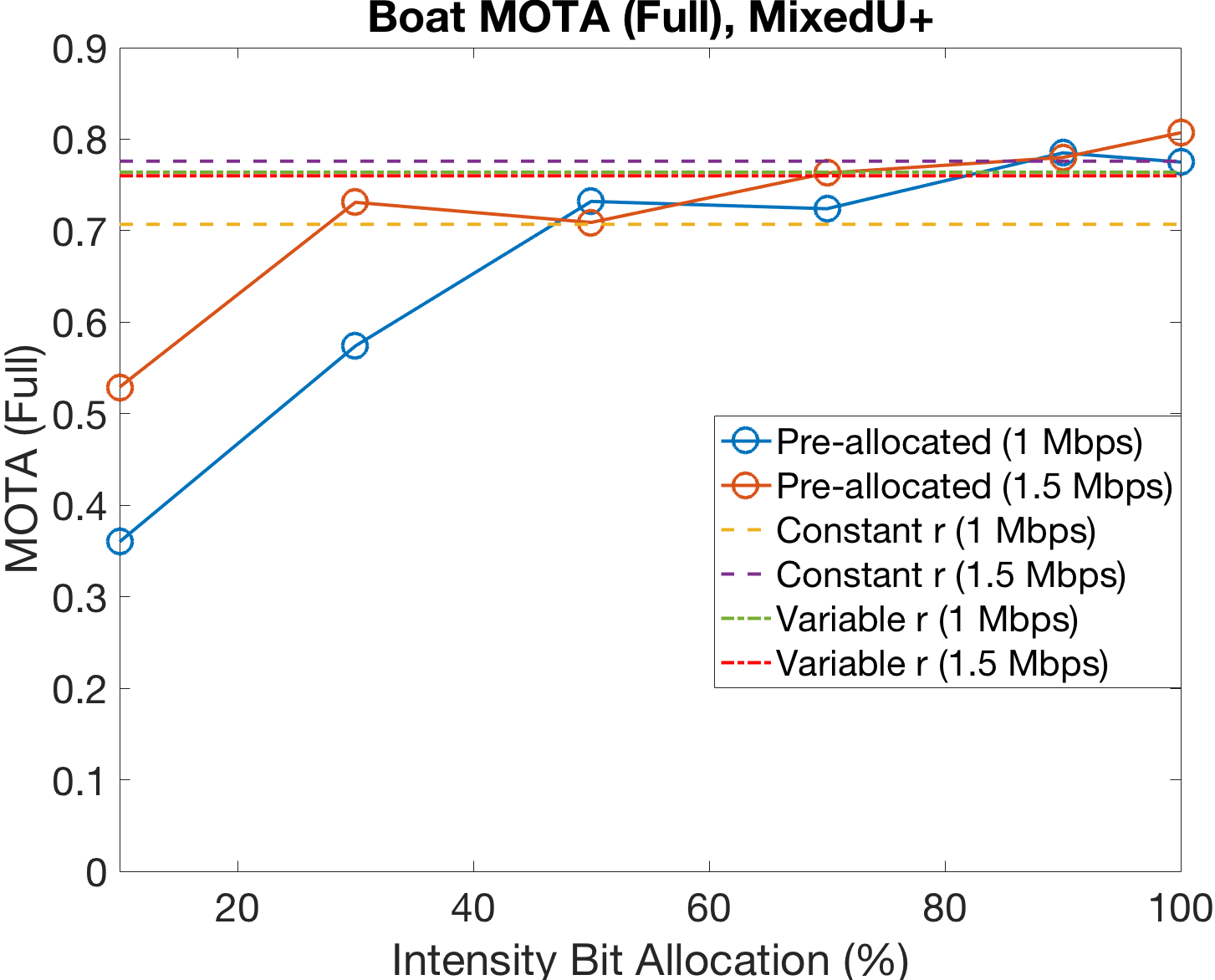}\par\caption*{(b) Watercraft}
    \includegraphics[width = \linewidth, height=4.5cm]{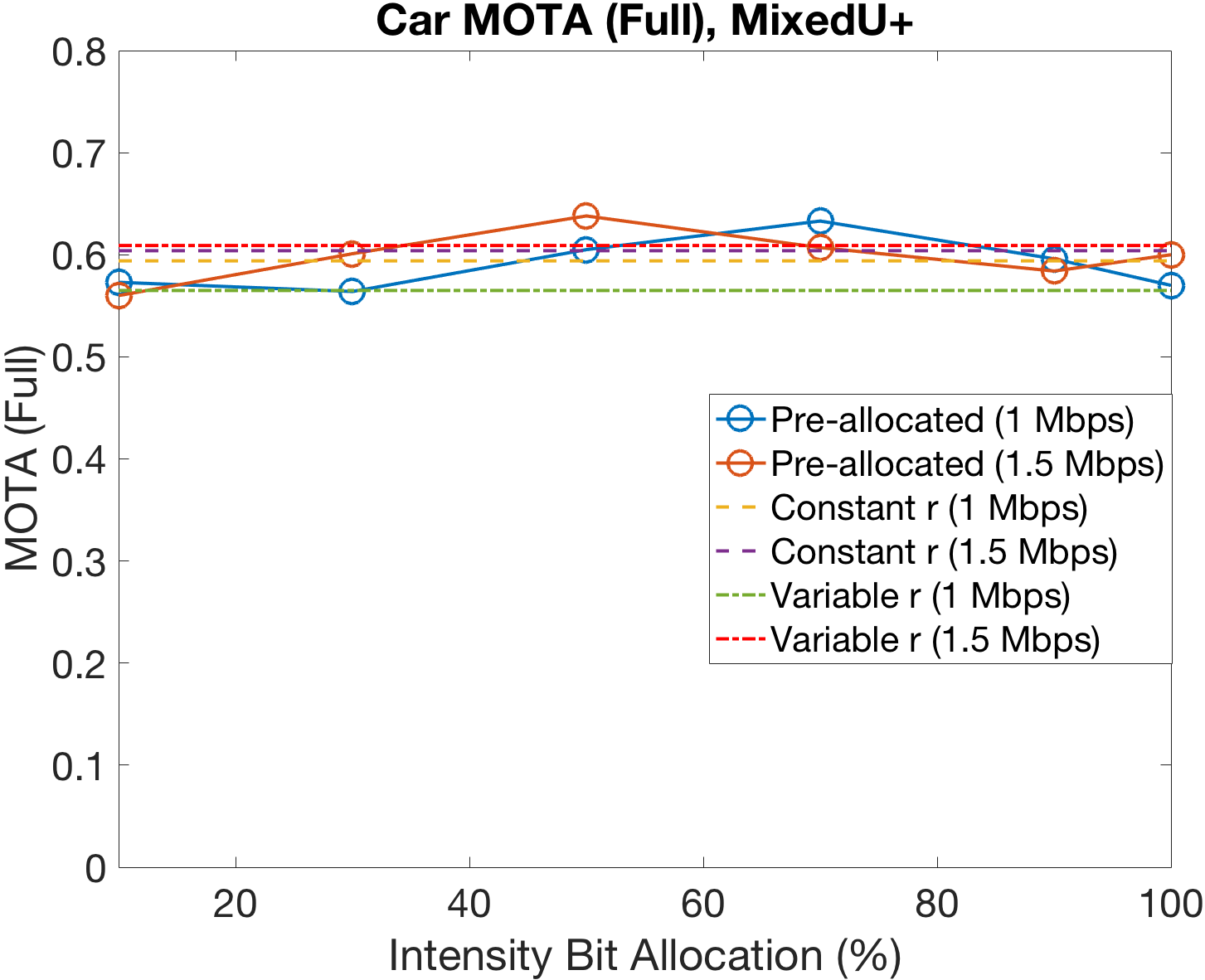}\par\caption*{(c) Car}
\end{multicols}
\caption{MOTA versus intensity bit rate allocation (prefixed allocation and jointly optimized with constant and variable $r$.)}
\label{MOTA_bit_rate_comp}
\end{figure*}
\begin{figure*}[t]
\begin{multicols}{2}
    \noindent
    \includegraphics[width = 0.75\linewidth, height=4.5cm]{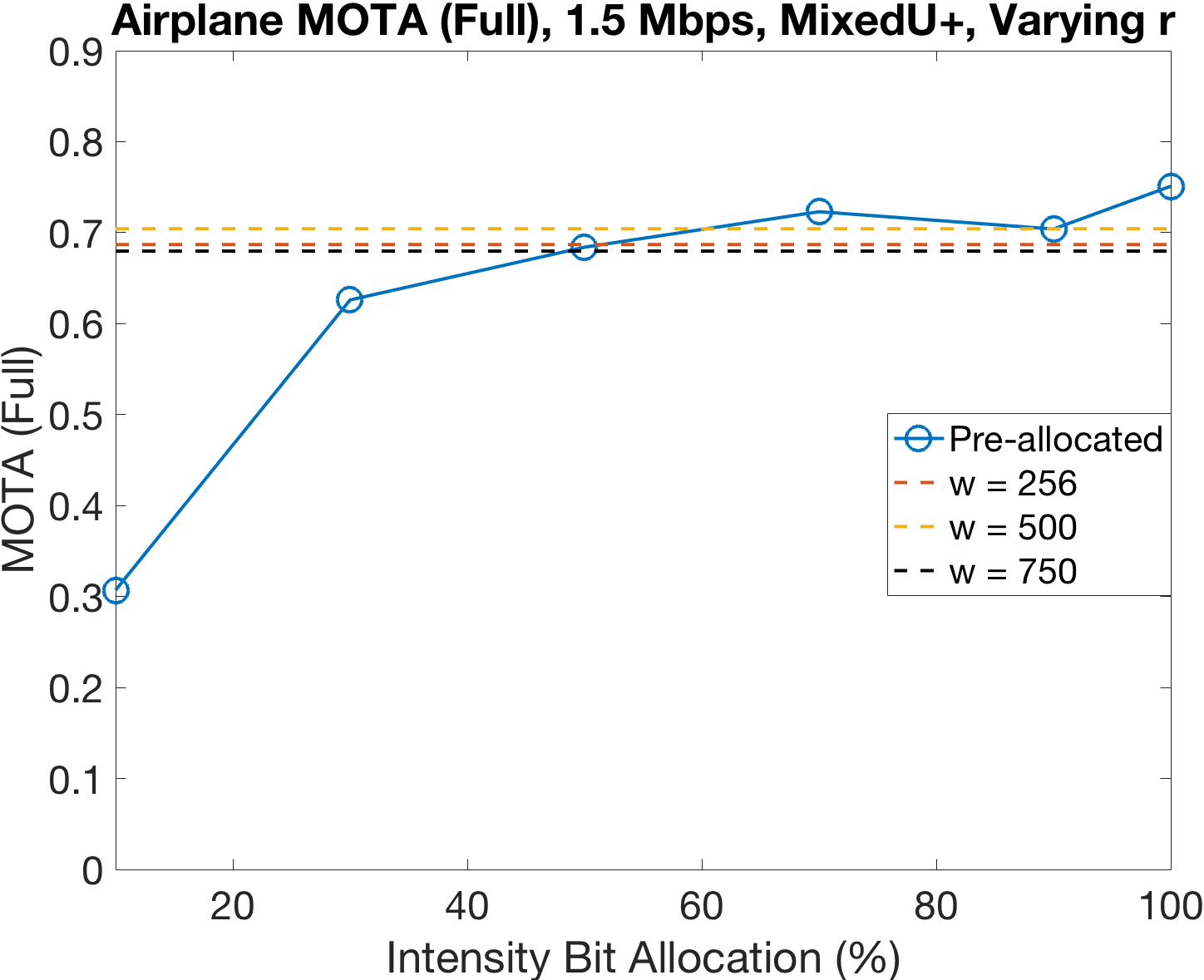}\par\caption*{(a) Airplane ($1.5$ Mbps)}
    \includegraphics[width = 0.75\linewidth, height=4.5cm]{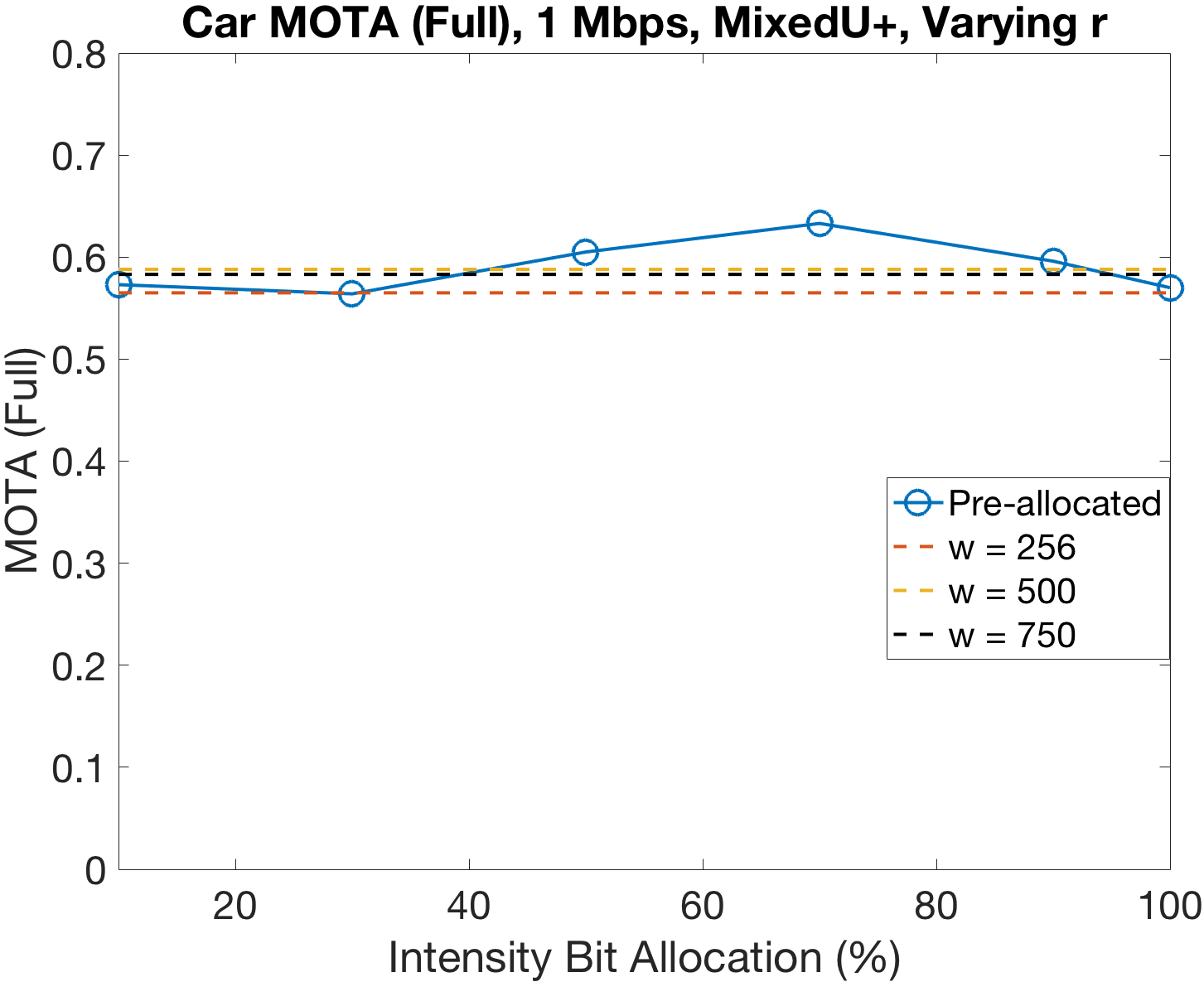}\par\caption*{(b) Car ($1$ Mbps)}
\end{multicols}
\caption{MOTA versus intensity bit rate for varying $w_e$ (prefixed allocation and jointly optimized with variable $r$).}
\label{MOTA_bit_rate_w}
\end{figure*}
\subsection{Performance with Edge Enhancement at different bit rates}
\label{Performance_edge_enhancement_event_comp}
It is clearly seen from Section \ref{Performance_edge_enhancement} that edge enhancement improves the performance of the system especially for higher distortions. In this section, we compare the performance of the system at different bit rates, with Pristine and MixedU+ object detectors, which are trained with no distortion and using system generated distortions in a 2-step method respectively, as described in \cite{ReImagine_Sys_1}. As in Section \ref{Performance_edge_enhancement}, we use the events only for edge reconstruction without object detection on the events and fusion network. We compare the MOTA metric for different bit rates as shown in Fig. \ref{fig_edge_rec_bit_Rate}. The $\%$ bit rate allocation of the intensity modality out of the total bit rate is varied from $10 \%$ to $100 \%$, with the rest of the bit rate allocated for the events. The MOTA metric improves with the $\%$ bit rate being allocated to the intensity modality, thereby implying better performance of the system towards higher intensity bit rates. Additionally, the system performance is better with MixedU+ detector compared to the Pristine detector for almost all experimental test cases. Moreover, with increasing bit rate to $1.5$ Mbps, system performance improves, especially at lower $\%$ allocated bit rates to the intensity modality, implying better system performance with more bits. For the airplane sequence, additional experiments have been carried out with $0.6$ Mbps which confirm the trend. In the subsequent sections, the experiments are done with the MixedU+ object detector.
\begin{figure*}[t]
\begin{multicols}{2}
    \noindent
    \includegraphics[width = 0.75\linewidth, height=4.5cm]{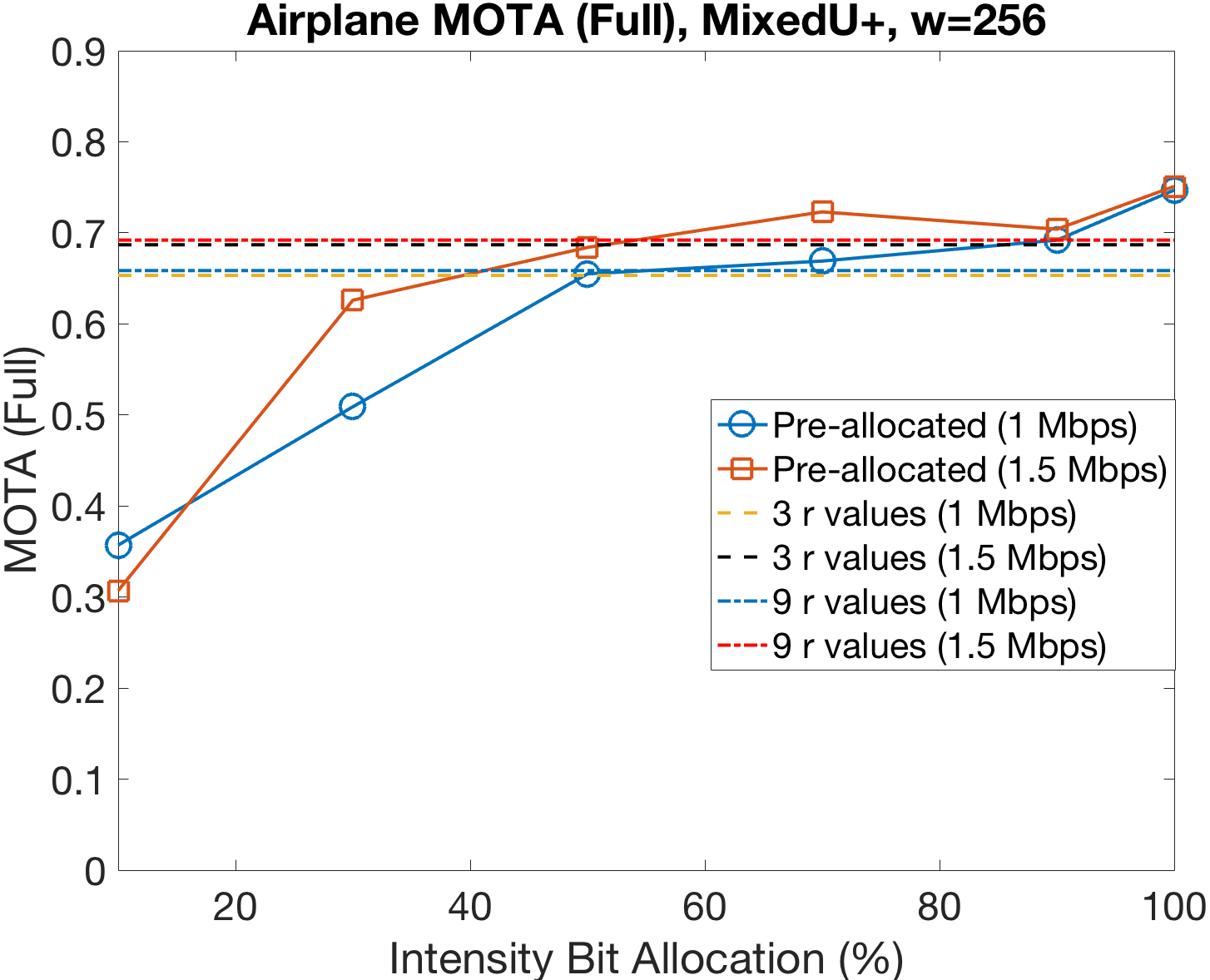}\par\caption*{(a) Airplane}
    \includegraphics[width = 0.75\linewidth, height=4.5cm]{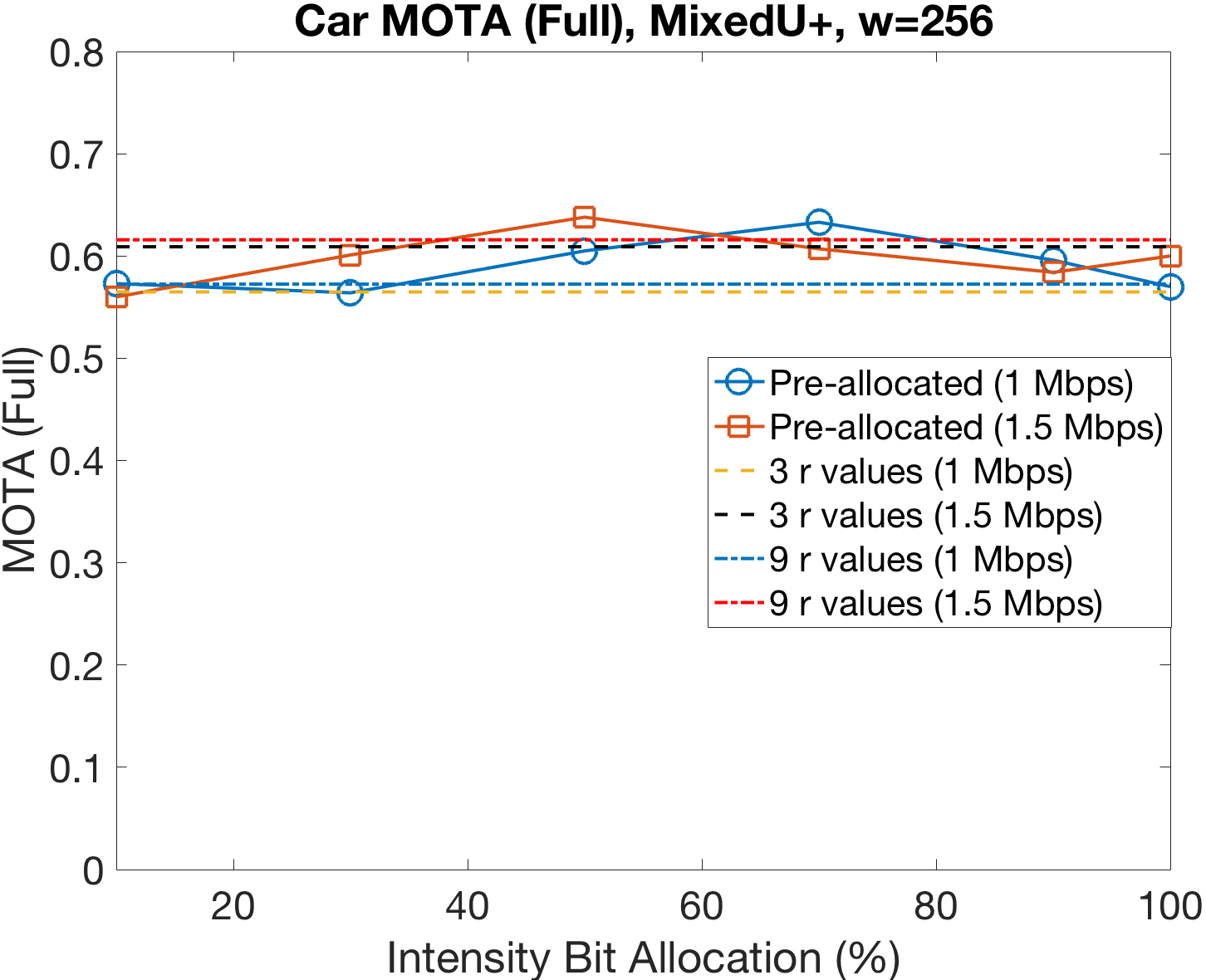}\par\caption*{(b) Car}
\end{multicols}
\caption{MOTA versus intensity bit rate allocation for variable $r$. $w_{e} = 256$ is used for all these experiments.}
\label{MOTA_bit_rate_r}
\end{figure*}
\subsection{Performance with fusion network}
\label{Performance_fusion_network}
It this section we evaluate the system performance of the events in object detection and tracking along with the fusion network in the system. The events are used to not only perform edge reconstruction as in Sections \ref{Performance_edge_enhancement} and \ref{Performance_edge_enhancement_event_comp}, but also to detect and track objects from the event frames and fuse the information from the event and intensity modalities. In Fig. \ref{fig_fusion}, the Original Fusion Model implies that the events are used only for edge enhancement, while the Updated Fusion Model implies that the events are used for edge enhancement, object detection, tracking and finally fusion. It is observed in Fig. \ref{fig_fusion} that the updated fusion network helps in improving the MOTA tracking performance of the system for all the allocated intensity bit rates as a fraction of the total bit rate of $1.5$ Mbps. MixedU+ object detector \cite{ReImagine_Sys_1} has been used in these experiments. It clearly shows that the events in the system help in improving the MOTA performance metric.
\begin{figure*}[t]
\begin{multicols}{3}
    \noindent
    \includegraphics[width = \linewidth, height=4.5cm]{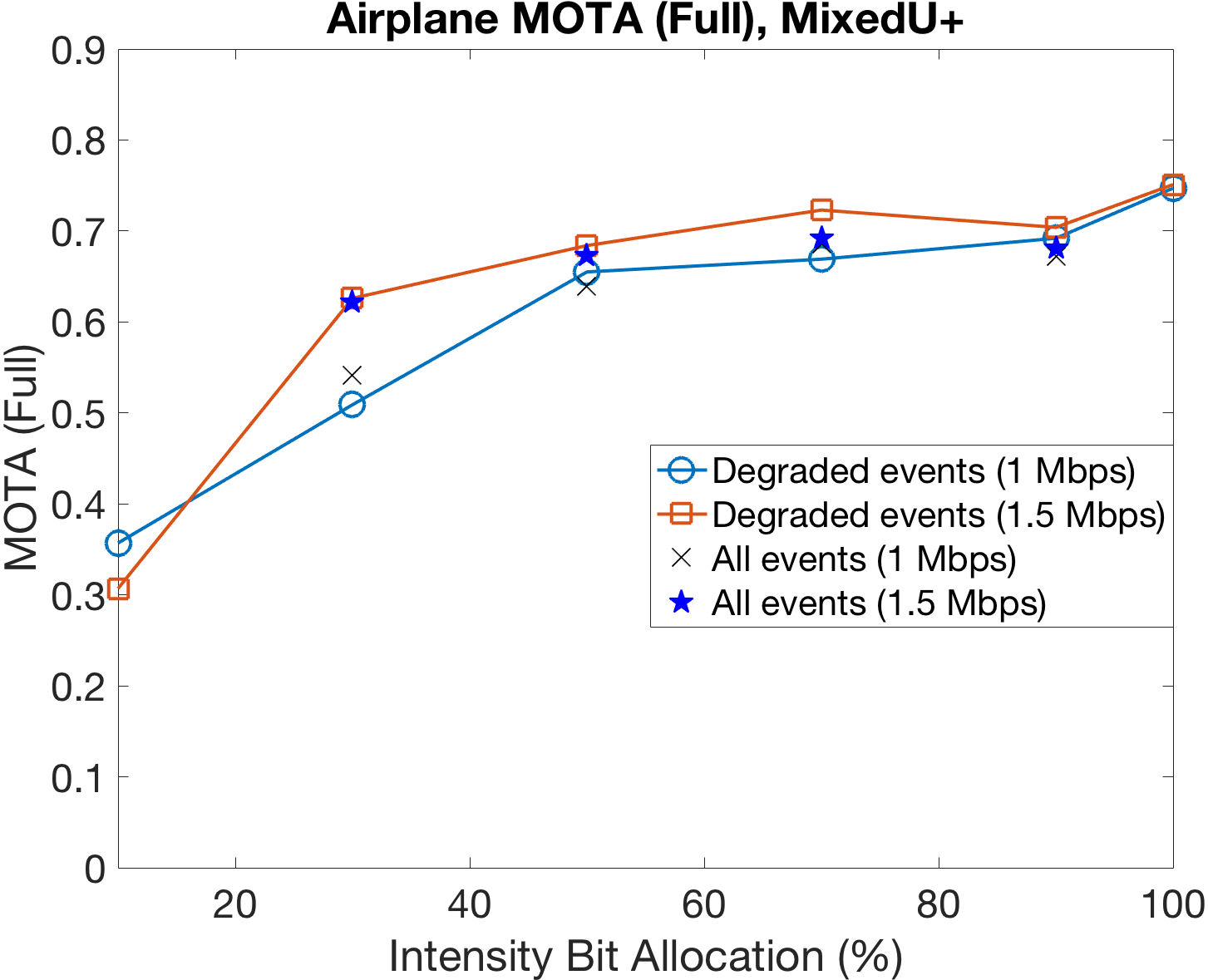}\par\caption*{(a) Airplane}
    \includegraphics[width = \linewidth, height=4.5cm]{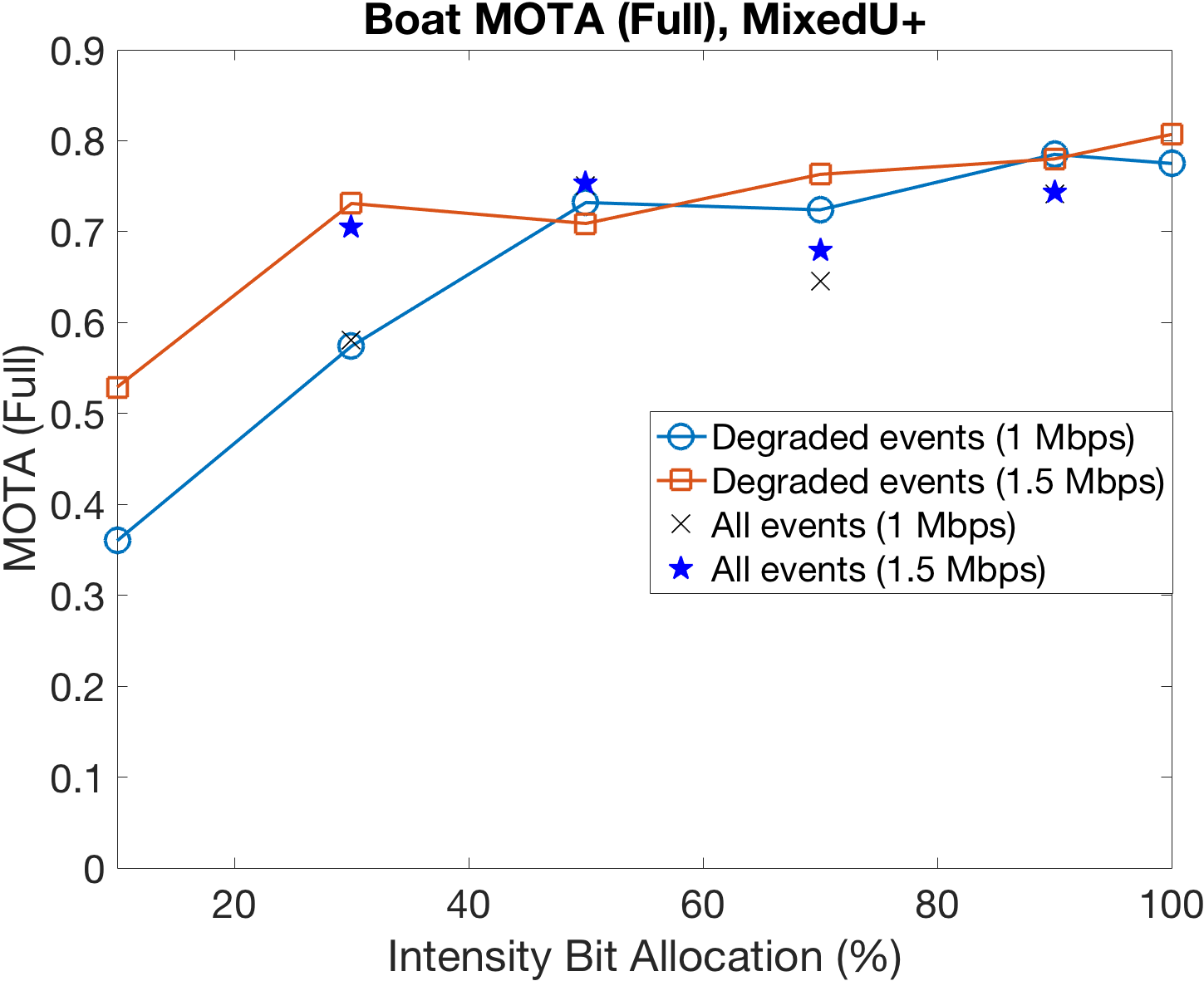}\par\caption*{(b) Watercraft}
    \includegraphics[width = \linewidth, height=4.5cm]{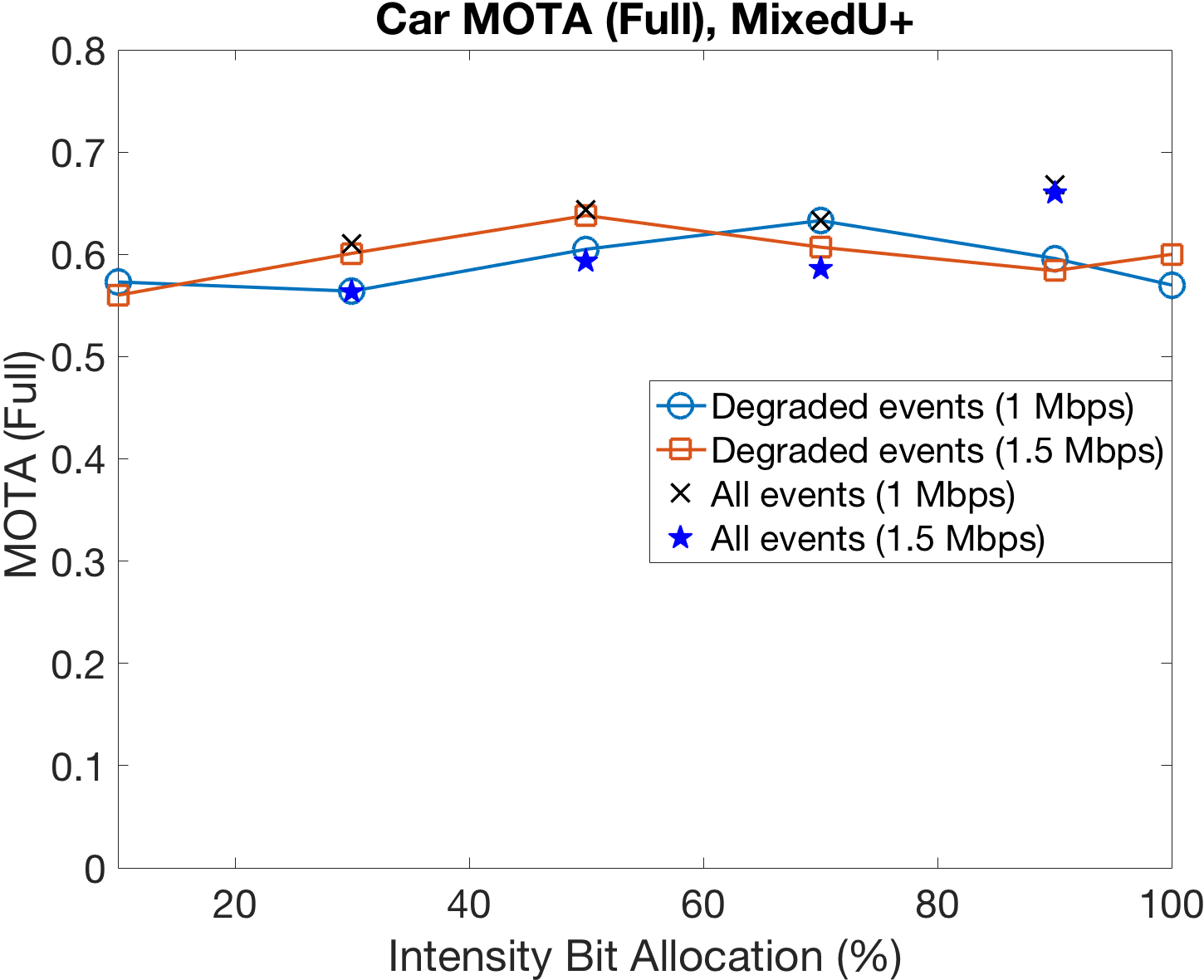}\par\caption*{(c) Car}
\end{multicols}
\caption{MOTA versus intensity bit rate allocation for original and distorted events.}
\label{MOTA_org_dist_events}
\end{figure*}

\subsection{Performance comparison with joint distortion optimization}
\label{Sys_per_joint_dist_opt}
In this section, we compare the performance of the system with prefixed bit allocation for a particular bit rate and the joint optimal allocation of the bits for intensity and event modalities.

\subsubsection{Variation of Bit Rates per frame}
\label{var_bit_rate_per_frame}
The variation of the fraction of the allocated bit rate for events out of the total bit rate over the frames is shown in Fig. \ref{fig_bit_rate_comp}. We compare the allocated fraction of bits for the events for two cases: (a) constant $r$, and, (b) variable $r$ in Eqn. \ref{eqn:eqlabel4}. For the constant $r$ case, $r_{4} = 1$, while for the variable $r$ case, $r_{4} \in \{1, 2, 3\}$. For both the constant and variable $r$ cases, $r_{8} = 2r_{4}$, $r_{16} = 4r_{4}$, $r_{32} = 4r_{4}$, and so on for higher sized QT blocks. It is seen that the fraction of the bits allocated to the events is lower for the variable $r$ case than the constant $r$ case (shown as $r=var$ and $r=const$ in Fig. \ref{fig_bit_rate_comp}). This is primarily due to the fact that for the variable $r$ case, the dynamic optimization can optimize over different $r$, hence providing better optimized bit rates.

\subsubsection{Performance variation}
\label{Sys_per_var}
Performance of the system in terms of MOTA is computed for pre-fixed allocated fraction of bit rate for the intensity out of the total bit rate, and compared with the MOTA for the system jointly optimized for the various bit rates using Eqn. \ref{eqn:eqlabel4}. Both a constant and variable $r$ are used in the joint optimization of the Intenisty-Event modalities. The results are shown in Fig. \ref{MOTA_bit_rate_comp}. Results of the jointly optimized system are shown as dashed lines, while the MOTA for the pre-fixed allocated bit rate is shown as solid lines. 

Pre-fixed allocation of the bits between the intensity and event modalities are able to achieve slightly better MOTA than jointly optimized intensity and event bit rates. For the airplane sequence, the best MOTA is achieved for intensity:event bit rate ratio of 100:0, while for the car and boat sequences, the best MOTA is achieved for some other intensity:event bit rate ratio. The MOTA plots for the jointly optimized intensity-event modalities indicate that the system performance can reach close to the highest MOTA value (corresponding to the pre-fixed bit allocation). Moreover, the MOTA for the variable $r$ cases has marginally higher values than for the fixed $r$ cases for most of the experimental results. This is primarily due to the fact that the variable $r$ has better optimized intensity-event split owing to relaxation in the optimization parameter space. Additionally, we can clearly see the trade-off between computational requirements and MOTA from the plots in Fig. \ref{MOTA_bit_rate_comp}. While the pre-fixed allocation of bits for the best MOTA has to be obtained by computing different intensity:events allocation ratios, for instance $50:50, 70:30, 90:10, 30:70, 10:90$, in the joint optimization algorithm, the computation for allocating the bits has to be performed only once. Still, the jointly optimized intensity-events provide us with MOTA which is close to the best possible MOTA.

\subsection{Performance Comparison with Varying Weights for Event Distortion}
In this section, we show the results of experiments in which we vary the weight of the distortion, $w_e$ in Eqn. \ref{eqn:eqlabel12}, for $w_{e} = 256, 500, 750$ for variable $r$ ($r=var$) as described in Section \ref{Sys_per_var}. Figure \ref{MOTA_bit_rate_w} (a) and (b) shows the MOTA performance for the airplane sequence at $1.5$ Mbps, and car sequence at $1$ Mbps respectively. For $w_{e} = 500$, the MOTA performance metric of the system is marginally better than for other $w_e$ values for most of the cases, as shown in Fig. \ref{MOTA_bit_rate_w}. Similar improvement in MOTA is seen for these sequences at other bit rates as well. However, the MOTA improvement is not significant as it basically implies one or two fewer false positives. For instance, in the car sequence at $1$ Mbps, the MOTA is $0.5883$ at $w_{e} = 500$, implying $2$ fewer positives compared to $w_{e} = 750$ for the same sequence with $107$ frames. For this sequence each false positive increases or decreases the MOTA by $0.0025$.    
\subsection{System performance with Different Event Compressions}
In this section, we perform experiments by varying the search space of the parameter $r$, which is the PDR for event encoding in the joint intensity-event optimization algorithm. We compare the system performance by keeping $r$ constant, and additionally varying $r$ over $3$ and $9$ values in order to find out the optimal choice of $r$ for maximizing MOTA. From Fig. \ref{MOTA_bit_rate_r}, it is evident that the improvements in MOTA by increasing r from $3$ to $9$ is marginal for both the airplane and car sequences at $1$ and $1.5$ Mbps. Although, we increase the PDR $r$ values such that a higher $r$ would remove more events near the vicinity of each event, we constraint the PDR from removing events in the adjacent blocks of the QT.  
\subsection{Performance with Degraded and Actual Events}
In this section we test the system performance with original and compressed events. The intensity bit rate is varied as a fraction of the total bit rate with the events used as original and compressed as shown in Fig. \ref{MOTA_org_dist_events}. For the compressed events, the bit rate is the remaining $\%$ of the bit rate after allocating the desired bit rate for intensity. For instance, when $70 \%$ bit rate is allocated to intensity, the compressed event bit rate is $30 \%$. For majority of cases, the MOTA performance of the system for compressed events is better or the same as that of the performance with original undistorted events. It must be noted that the event based processing networks: the edge enhancement network and the event object detector have been trained on original events only. However, the results, indicate that training these networks with compressed events is not essential from the system performance point of view. This is an interesting observation. We reason that this behavior is due to the fact that the interpolated ILSVRC VID frames have both small and large motion of the objects and scene, which results in generation of dense and sparse events using ESIM \cite{rebecq2018esim}. The compression step is in a way a sparsifying step, which results in sparse events with similar characteristics to the events generated from the interpolated ILSVRC VID dataset.

\subsection{Experiments with blurry sequences}
\label{blurry_seq}
\begin{figure}[htbp]
\centerline{\includegraphics [width=0.8\linewidth]{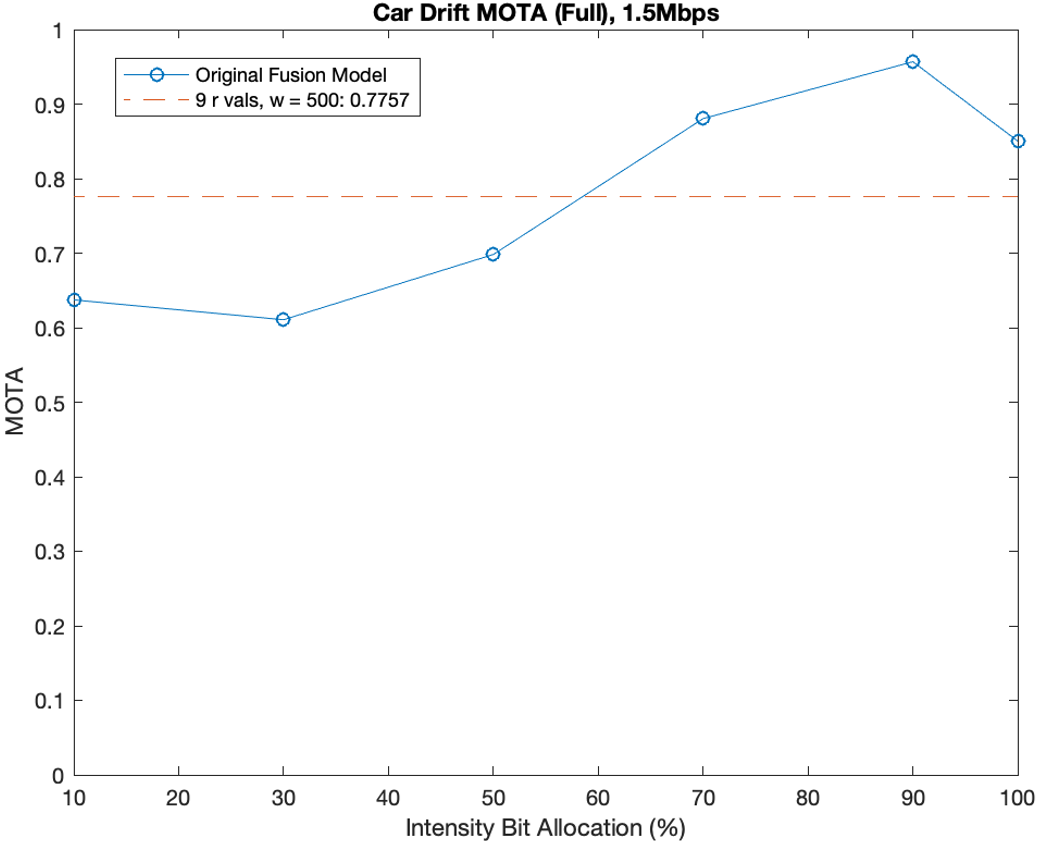}}
\caption{MOTA versus intensity rate for blurred sequence.}
\label{MOTA_NFS_data_results}
\end{figure}

It is evident in literature such as \cite{lin2020learning}, \cite{zhang2020hybrid} that events are quite essential during processing of blurred and/or high speed motion. We perform experiments with blurred sequences using the car$\_$drifting sequence from publicly available Need for Speed (NFS) dataset \cite{NFS_dataset}, which contains frames at $240$ fps and motion blurred frames at $30$ fps. In order to generate events, we first interpolate frames to $960$ fps using Super SloMo \cite{jiang2018super}, while we generate events from these frames using ESIM \cite{rebecq2018esim}. The blurred frames and events are used for testing. The maximum MOTA for the system is at intensity $:$ event bit rate allocation of $90 : 10$ of the bit rate at $1.5$ Mbps as shown in Fig. \ref{MOTA_NFS_data_results}. The jointly optimized intensity-event modality has been done with variable $r$ over $9$ possible candidate values with event distortion weight $w_{e} = 500$. The MOTA for the jointly optimized intensity-event is approximately $80\%$ of the maximum MOTA achieved with pre-fixed allocated bit rate.
\section{Discussion}
The MOTA metric primarily indicates the tracking performance of this system. We have shown the system performance in terms of MOTA with and without events. The improvement of MOTA is dependent on the sequence and the operating bit rate. For instance, at low bit rates the events contribute more towards improving the MOTA score. On the other hand, at high bit rates, events do not seem to make a difference. For blurry sequences, the events contribute in improving the MOTA score. Additionally, it is observed that with pre-fixed allocated intensity and event bit rates, the MOTA performance of the system reaches its maximum value. In our experiments, the joint allocation of bits based on distortion between the intensity and event modalities provide us with MOTA score which is close to the maximum value but still does not provide the highest possible MOTA value. This is due to the fact that we jointly optimize for the rate-distortion trade-off between the intensity and event modalities and not directly between the rate and MOTA trade-off. The experiments do not establish a linear relation between distortion and MOTA values. Higher distortion does not necessarily indicate a lower MOTA score. Secondly, due to the lower computational power available on the chip, the MOTA metric cannot be computed on the chip, and therefore we resort to computing rate-distortion optimization on it. Additionally, the camera motion in the scene generates several background events which can lead to the joint optimization of the intensity-event to converge to a different point than without scene motion. This is particularly evident in the car$\_$drifting sequence, as shown in Section \ref{blurry_seq}.

Nevertheless, despite these shortcomings, the MOTA score from jointly optimized event-intensity bit rate is close to the maximum MOTA obtained from the pre-fixed bit rate allocation. The joint optimization allows for the use of a less computationally powerful chip compared to the one required for generating distorted data with different pre-fixed intensity:event ratios.

\section{Conclusions}
This paper proposes a novel tracking algorithm using the intensity and event modalities for a computationally constrained chip device. A novel intensity - event joint dynamic optimization framework is proposed which minimizes the rate-distortion equation jointly. The distorted events and intensity frames are used on the host for object tracking which is evaluated using the MOTA metric. Tracking and detection of objects are done in the intensity and event mode separately before they are fused together in a late fusion strategy. The events are also used for edge enhancement of the distorted intensity frames. The paper provides an in-depth study of the different system parameters, ablation studies, and comparison of joint bit allocation and pre-fixed allocation of intensity and event modalities.

\ifCLASSOPTIONcaptionsoff
  \newpage
\fi



%\bibliographystyle{IEEEtran}
%\begin{thebibliography}{1}

\bibliography{bibtex/bib/IEEEabrv.bib,bibtex/bib/IEEEexample.bib}{}
\bibliographystyle{IEEEtran}


%\end{thebibliography}


\begin{comment}
\begin{IEEEbiography}{Srutarshi Banerjee}
Biography text here.
\end{IEEEbiography}

\begin{IEEEbiography}{Henry H. Chopp}
Biography text here.
\end{IEEEbiography}

\begin{IEEEbiography}{Jianping Zhang}
Biography text here.
\end{IEEEbiography}

\begin{IEEEbiography}{Zihao W. Wang}
Biography text here.
\end{IEEEbiography}

\begin{IEEEbiography}{Oliver Cossairt}
Biography text here.
\end{IEEEbiography}

\begin{IEEEbiography}{Aggelos Katsaggelos}
Biography text here.
\end{IEEEbiography}
\end{comment}

\end{document}